\documentclass[a4paper,fleqn,usenatbib]{mnras}

\usepackage{newtxtext,newtxmath}

\usepackage[T1]{fontenc}
\usepackage{ae,aecompl}

\usepackage{graphicx}	
\usepackage{amsmath}	
\usepackage{amssymb}	
\usepackage{pdflscape}  
\usepackage{subfig}     

\newcommand{\msun}{\ensuremath{M_{\odot}}}
\newcommand{\rsun}{\ensuremath{R_{\odot}}}
\newcommand{\taun}{\ensuremath{\tau_{\rm n}}}
\newcommand{\tauni}{\ensuremath{\tau_{\rm n,i}}}
\newcommand{\taunx}{\ensuremath{\tau_{\rm n,x}}}
\newcommand{\taugw}{\ensuremath{\tau_{\rm GW}}}
\newcommand{\taugwstar}{\ensuremath{\tau_{\rm GW}^\star}}
\newcommand{\taud}{\ensuremath{\tau_{\rm d}}}
\newcommand{\afai}{\ensuremath{A_{\rm f}/A_{\rm i}}}
\newcommand{\ain}{\ensuremath{A_{\rm i,min}}}
\newcommand{\aix}{\ensuremath{A_{\rm i,max}}}
\newcommand{\ai}{\ensuremath{A_{\rm i}}}
\newcommand{\af}{\ensuremath{A_{\rm f}}}
\newcommand{\astar}{\ensuremath{A^\star}}
\newcommand{\emin}{\ensuremath{e_{\rm min}}}
\newcommand{\emax}{\ensuremath{e_{\rm max}}}
\newcommand{\mdn}{\ensuremath{M_{\rm DN}}}
\newcommand{\mdnzero}{\ensuremath{M_{\rm DN0}}}
\newcommand{\msec}{\ensuremath{m_2}}
\newcommand{\mpri}{\ensuremath{m_1}}
\newcommand{\mprin}{\ensuremath{m_{1,\rm min}}}
\newcommand{\mprix}{\ensuremath{m_{1,\rm max}}}
\newcommand{\fkn}{\ensuremath{f_{\rm kn}}}
\newcommand{\kkn}{\ensuremath{k_{\rm kn}}}
\newcommand{\Rkn}{\ensuremath{R_{\rm kn}}}
\newcommand{\prompt}{\textit{prompt}}
\newcommand{\delayed}{\textit{delayed}}
\newcommand{\Fp}{\ensuremath{F_{\rm p}}}

\newcommand{\kalpha}{\ensuremath{k_\alpha}}
\newcommand{\alphamns}{\ensuremath{\alpha_{\rm MNS}}}

\title[Rate of merging DNS]{On the delay times of merging double neutron stars}
\author[L. Greggio et al.]{
Laura Greggio,$^{1}$\thanks{E-mail: laura.greggio@inaf.it}
Paolo Simonetti,$^{2,3}$
and Francesca Matteucci$^{2,3}$
\\
$^{1}$INAF, Osservatorio Astronomico di Padova, Vicolo dell'Osservatorio 5, I-35122 Padova, Italy \\
$^{2}$INAF, Osservatorio Astronomico di Trieste, Via G.B. Tiepolo 11, I-34143 Trieste, Italy\\
$^{3}$Dipartimento di Fisica, Sezione di Astronomia, Universit\'a degli Studi di Trieste, Via G.B. Tiepolo 11, I-34143 Trieste, Italy
}

\date{Accepted XXX. Received YYY; in original form ZZZ}

\pubyear{2020}

\begin{document}
\label{firstpage}
\pagerange{\pageref{firstpage}--\pageref{lastpage}}
\maketitle

\begin{abstract}
The merging rate of double neutron stars (DNS) has a great impact on many astrophysical issues, including the interpretation of gravitational waves signals, of the short Gamma Ray Bursts (GRBs), and of the chemical properties of stars in galaxies. Such rate depends on 
 the distribution of the delay times (DDT) of the merging events. In this paper we derive a theoretical DDT of merging DNS following from the characteristics of the clock controlling their evolution. We show that the shape of the DDT is  governed by a few key parameters, primarily the lower limit and the slope of the distribution of the separation of the DNS systems at birth. With a parametric approach we investigate on the observational constraints on the DDT from the cosmic rate of short GRBs and the Europium to Iron ratio in Milky Way stars, taken as tracer of the products of the explosion. We find that the local rate of DNS merging requires that $\sim 1 \%$ of neutron stars progenitors live in binary systems which end their evolution as merging DNS within a Hubble time. The redshift distribution of short GRBs does not yet provide a strong constraint on the shape of the DDT, although the best fitting models have a shallow DDT. The chemical pattern in Milky Way stars requires an additional source of Europium besides the products from merging DNS, which weakens the related  requirement on the DDT. At  present both constraints can be matched with the same DDT for merging DNS. 

\end{abstract}





\section{Introduction}

The detection of gravitational waves in the most recent years has opened a new venue for astrophysics, in particular with respect to the topic of the merging of compact objects.  So far, 11 confirmed events  of binary Black Hole merging, and 2 confirmed binary neutron star merging have been collected
\citep{abbott19,collaboration2020gw190412,abbott20}. 
For more than 3 decades the merging of two neutron stars has been proposed to be at the origin of the short Gamma Ray Bursts (SGRB) phenomenon \citep{paczynski86,eichler89,narayan92,Li98,tutukov07,giacomazzo13}. These events have also been proposed as responsible of part of the enrichment of  some r-process elements, like Europium  \citep{lattimer77,meyer89,freiburg99,rosswog99,korob12,hoto13}. This picture has gained strong credit with the observations of the counterpart of the gravitational waves detection GW170817 \citep[see,e.g.][]{ciolfi20}, originated from the merging of two neutron stars \citep{abbott17a}, in gamma-rays \citep{abbott17b}, X-rays and optical wavelengths \citep{abbott17c}. The latter transients correspond to the electromagnetic emission which accompanies an explosive event resulting from the merging of the two neutron stars, often addressed to as \textit{kilonova} \citep[see, e.g.][]{tanvir13,berger14}. Therefore, the rate of merging of binary neutron stars in stellar systems is a fundamental element to interpret the data measured by gravitational waves detectors, to model the occurrence of SGRBs in galaxies and as a function of redshift, as well as to model the chemical evolution of galaxies, especially for what concerns those nucleosynthetic products from kilonovae.

In a stellar system the merging rate of double neutron stars (hereafter DNS) results from the convolution of the star formation history with the distribution of the delay times \footnote{The delay time is the time elapsed between the birth of the binary system and its final merging.} (herafter DDT) of the events. If the delay times are short the rate of DNS merging closely follows the star formation rate, similar to the case of core collapse supernovae (CC-SNe). The fact that some SGRBs have been associated with early type galaxies \citep{fong13} argues for some events occurring well after star formation has ceased, i.e. for a component with long delay times, similar to the case of SNe Ia. In other words, the DDT of DNS merging is likely to be a wide distribution function, including \prompt\ as well as \delayed\ events. The DDT, which is proportional to the DNS merging rate from one single stellar generation, is a crucial ingredient for modelling the occurrence of kilonovae explosions in galaxies of different types, and, in turn, for calculating the timescale over which their products are released to the interstellar medium. 

The redshift distribution of SGRBs and their properties (e.g. luminosity, fluence, duration) have been used to characterize the DDT of DNS merging events in several works  \citep[e.g.][]{guetta06,virgili11,davanzo14,wanderman15}. These efforts combine a description of the cosmic star formation history with parametrized functions for the DDT, to construct models and compare them to the data. Often the results indicate a DDT which scales with the inverse of the delay time \citep[see also][]{ghirlanda16}. 

Abundance and abundance ratios of Milky Way stars can also be used to derive important constraints on the DDT of merging DNS. Models of chemical evolution \citep{matteucci14} including merging DNS as producers of a pure r-process element like Europium (Eu) have shown that these events can be responsible for the total Eu production in the solar vicinity, but only if the delay time for merging is constant and quite short (1 Myr from the formation of the system), and if the mass range of neutron stars progenitors extends from 9 to 50 \msun.
Alternatively, if both CC-SNe and merging DNS contribute to the Eu production, the
data could be reproduced allowing longer delay times for the merging DNS. \citet{cote18} pointed out that in order to reproduce the [Eu/Fe] vs. [Fe/H] relation in the solar vicinity, which is very similar to that of any $\alpha$-element relative to Fe, a DDT  such as that of SNe Ia would produce results at variance with the observations. They tested time delay distribution functions scaling as $\propto t^{-a}$ (with $a=1$ and 1.5).
Adopting a modified power law for the DDT, and exploring a variety of options, \citet{simonetti19} showed that in order to reproduce the observed cosmic rate of SGRBs and to justify the occurrence of the GW170817 event in an early type galaxy, the DDT should be rather shallow, with an average coalescence time of 300-500 Myr. On the other hand, the evolution of [Eu/Fe] vs [Fe/H] abundance ratio in the solar vicinity requires a shorter timescale for the Eu production, which could be accomplished either adding an early and continuous contribution by CC-SNe, or assuming that the fraction of merging DNS per unit mass of the parent stellar population varies with time, with more events in the past. 

The DDT of merging DNS can also be computed numerically with the Binary Population Synthesis (BPS) technique, which takes full advantage of the results of the stellar evolution theory \citep[e.g.][]{tutukov93,nelemans01,dominik12,mennekens16,mapelli17,giacobbo18,eldridge19,tang20}.
The typical evolutionary path leading to the formation of a DNS system which merges within a Hubble time starts with a close binary made of two massive stars, progenitors of a neutron star remnant \citep{faber12,tauris17}. The primary evolves and upon expansion it may overfill its Roche Lobe and lose mass, which may or may not be accreted by the companion. In any case, the primary will eventually explode as a supernova leaving behind a neutron star remnant. Upon evolution, the secondary expands and fills its Roche Lobe. A Common Envelope (CE) phase follows, during which the binary system shrinks because of friction. Part of the orbital energy is transferred to the CE which is eventually dispersed in the interstellar medium, leaving behind a system composed of a neutron star and a Helium star companion. During its evolution the Helium star expands and may again fill its Roche Lobe, possibly leading to another CE phase, and further shrinking the system. The supernova explosion from the secondary will thus leave a close binary neutron star system, which will eventually merge because of angular momentum loss due to the emission of gravitational waves radiation. The DNS formation may however be aborted when either of the two supernova explosions occur, as the system may disrupt because of the effect of the supernova kick.

The BPS computations start from a population of primordial binaries and follow their evolution through the several mass exchange phases. Many recipes need to be implemented in BPS codes, including those describing the response of the two stars to Roche Lobe overflow, the supernova kick and its impact on the system \citep[e.g.][]{bray18}, the binding energy of the stellar envelope, the efficiency of the CE phase in shrinking the system, the initial-final mass relation, the radius evolution of the individual components, as well as the dependence on the chemical composition of the stellar evolutionary models. In addition, the distribution of the binaries in mass of the primary, mass ratio and separation add parameters to the BPS computations.
Some of these ingredients are robust, some are founded on empirical data, some are poorly known, e.g. the CE efficiency. Meanwhile, the computation of the DDT of merging DNS involves following the evolution of the binaries from an initial separation of several hundreds of \rsun, needed to avoid premature merging, down to a final separation of a few  \rsun\ or less, in order to ensure merging within a Hubble time, through an intermediate phase in which the separation could reach a few thousands of \rsun\ \citep[see][]{belczynski18}. This requires an accurate description of the various mass exchange phases. Besides, as noticed in \citet{martyna19}, only a small fraction of the theoretical binary population ends up in DNS merging within a Hubble time, some systems merging before the formation of the two neutron stars, some ending up with too long coalescence timescales, some other because of disruption when either of the two supernovae explode. Therefore, the description of the supernova kick and of the system response to it strongly impact on the BPS results \citep[see][]{giacobbo18}.
The local rates of kilonovae predicted by BPS computations and reported in \citet{martyna19} show a large variance, likely due to different recipes implemented in the codes \citep{tang20}.  

In this paper we present an alternative approach for determining the DDT of DNS mergers, similar to that developed in \citet{greggio05} for the rate of type Ia supernovae (SNe Ia), i.e. focussing on the properties of the clock which governs the merging events. Rather than following the evolution of individual systems, we look at the parameters which predominantly control the delay time and derive the DDT from the distribution of these parameters. Based on this approach,
we elaborate  parametrized models for the shape of the DDT which results from the properties of the clock; then we derive constraints on the shape of the DDT by comparing the redshift distribution of SGRBs to models obtained combining the theoretical DDTs with the cosmic star formation history in \citet{madau14}. The local rate of kilonovae estimated by \citet{abbott20} is used to calibrate the models, which allows us to evaluate the efficiency of kilonova production from a stellar population. Finally we test the models on their ability to account for the chemical trend of Europium and Iron abundances in the Milky Way. A similar exercise was presented in \citet{simonetti19} where we adopted a shape for the DDT based on generic arguments in accordance to the results in \citet{greggio05} for SNe Ia. Here we revisit the problem with a more rigorous determination of the DDT, exploring the effect of the distribution of the binary masses, eccentricity and separations. Our results include a variety of DDTs which cover a wide parameter space. 
We anticipate that the shape of the DDT turns out to crucially depend on the distribution of the separations of the binary neutron stars at birth and on its lower boundary. This offers a key to appraise the results of BPS codes with respect to the various recipes implemented to follow the close binaries evolution.

As a note of caution we remark that the models presented here are applicable only to binaries which evolve in isolation, while merging DNS can also be produced by dynamical processes which take place in dense environments, e.g. Globular Clusters \citep{lee_10}. The contribution of the dynamical to the total rate of merging DNS events is unclear: according to \citet{belczynski18}, in old stellar populations the current rate of DNS merging from the dynamical channel is $\sim$ 150 times lower than the rate from isolated binaries. On the other hand, models for the cosmic rate of DNS mergers by \citet{santoliquido20} show that the dynamical channel contributes $\sim$ 1/3 of all the local events. Our arguments are based on a delay time which does not include the time taken by the dynamical interaction to form the DNS system; therefore they are relevant for the contribution to the merging DNS events from binaries which evolve unperturbed by the environment. 

The paper is organized as follows: in Sect. 2 we describe the properties of the coalescence timescales, in Sect. 3 we show our models for their distribution, illustrating the dependence on the distribution of the total mass, of the separation and eccentricity of the DNS systems at birth. In Sect. 4 we derive model  distributions of the total delay times, which include the time necessary to produce the DNS system. In Sect. 5 we discuss the constraints on the DDT from the SGRBs redshift distribution and derive an estimate for the efficiency of DNS system production from a stellar population needed to account for the local rate of kilonovae. In Sect. 6 we compare chemical evolution models constructed with our model DDTs  to the Eu and Fe trend in Milky Way stars; in Sect. 7 we summarize  our results and draw some conclusions.

\section{The delay time}

For a binary system which evolves in isolation the time elapsed between the formation of the primordial system and the final merging is the sum of the evolutionary lifetime of the secondary component (\taun, hereafter referred to as the nuclear delay) and the time taken by the gravitational wave radiation to bring the components into contact (\taugw, hereafter referred to as GWR delay). The first timescale is a function of the initial mass of the star ($m$) and of the chemical composition. For the sake of simplicity we neglect the dependence on metallicity, and adopt the relation
\begin{equation}
    \log{m} = 0.49 (\log{\taun})^2 - 7.80 \log{\taun} + 31.88
    \label{eq_taun}
\end{equation}
with mass in solar units and time years. Eq. (\ref{eq_taun}) was  
derived from fitting the lifetimes of the neutron stars progenitors in the \citet{limongi06} models, complemented with the \citet{bertelli09} tracks, with solar metallicity.  We notice that, due to interaction, in a close binary the masses of the components may change, leading to a modification of the evolutionary lifetimes. For example, the mass of the secondary may increase leading to a shortening of the nuclear delay with respect to Eq. (\ref{eq_taun}). In our simple approach we also neglect the lifetime of the CE phase which is likely very short compared to other timescales involved \citep[e.g.][]{Igoshev20}. In general, Eq. (\ref{eq_taun}), which is appropriate for stars evolving in isolation, represents an approximation to the evolutionary lifetime of the secondary star in a close binary; yet it accounts for a basic trend of this component of the delay time related to the different masses of the secondaries in the progenitor systems.

\begin{figure}
\includegraphics[width=\columnwidth]{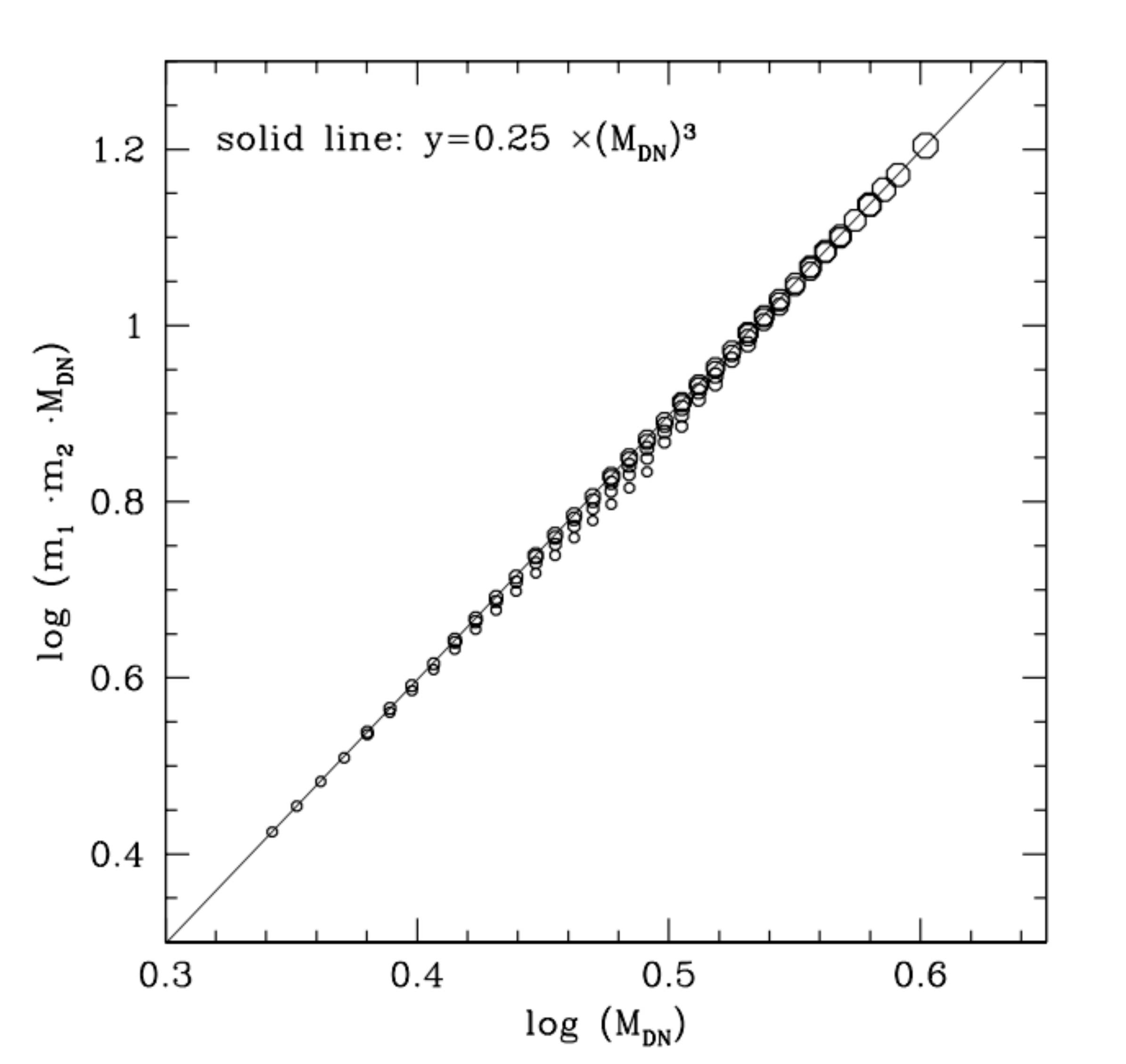}
\caption{The mass dependent term in Eq. (\ref{eq_taugw0}) is plotted as a function of the total mass of the binary, for $m_1$ and $m_2$ ($\leq m_1$) ranging from 1.1 to 2 \msun. Larger symbols refer to higher values of the primary mass. The solid line represents a good approximation to this term.}
\label{fig_fm}
\end{figure}

The time delay due to the action of the gravitational waves radiation can be expressed
as \citep{peters64}:
\begin{equation}
\taugw = \frac{0.15\,A^4}{m_1m_2(m_1+m_2)} \times (1 - e^2)^{7/2} \, {\rm Gyr}
\label{eq_taugw0}
\end{equation}
where $A$, $m_1$ and $m_2$ are respectively the separation and the masses of the components (in solar units), and $e$ is the eccentricity of the binary, all parameters evaluated at formation of the DNS system. 
Figure \ref{fig_fm} shows the mass dependent factor in Eq. (\ref{eq_taugw0}) as a function of the total mass of the binary ($\mdn = m_1 + m_2$), assuming that the mass of a neutron star varies between 1.1 \msun\ and 2 \msun, a range suggested by observational determinations \citep{martinez15, Antoniadis13}. It appears that the mass dependent factor can be well represented with the expression $y=0.25 \times \mdn^3$;
Eq. (\ref{eq_taugw0}) can then be approximated as
\begin{equation}
\taugw = \frac{0.6A^4}{\mdn^3} \times (1 - e^2)^{7/2} \, {\rm Gyr} .
\label{eq_taugw}
\end{equation}

\begin{figure}
\includegraphics[width=\columnwidth]{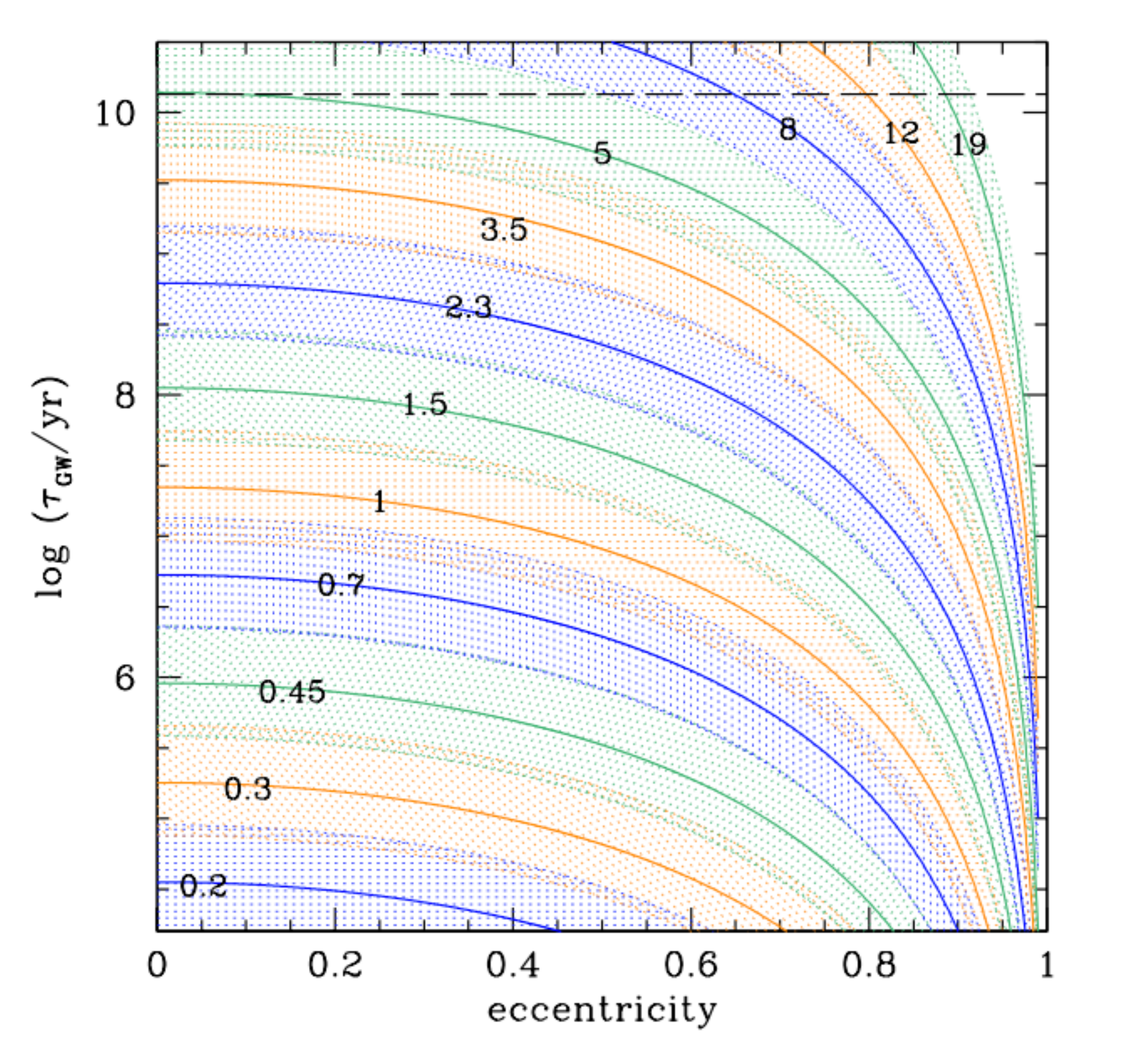}
\caption{The GWR delay as from Eq.(\ref{eq_taugw}), as a function of the eccentricity, for different values of the separation of the DNS system at formation ($A$) labelled on the solid lines (values are in units of \rsun)}. 
Solid lines adopt a total mass of the binary of 3 \msun, while the shaded areas show the range of \taugw\ for \mdn\ varying between 2.2 and 4 \msun\ at fixed separation and eccentricity. The dashed line is drawn at \taugw = 13.5 Gyr.
\label{fig_taugw}
\end{figure}

Figure \ref{fig_taugw} shows the GWR delay as a function of the eccentricity for different values of the separation and total mass of the binary. This figure illustrates the main characteristics of the GWR clock:
\begin{itemize}
\item \taugw\ is sensitive to all three parameters and it decreases as the eccentricity and the binary mass increase, and as the separation decreases;
\item although a value of \taugw\ corresponds to a variety of combinations of the parameters, there is a maximum delay achievable with a given value of the separation. In other words, all close systems merge on a short timescale, while only wide systems can merge on a long \taugw;
\item most of the relevant parameter space is limited to the range $A \lesssim 10 \rsun $, since wider systems can contribute to merging within a Hubble time only if born with very high eccentricities. Indeed, the range of eccentricities leading to merging timescales shorter than the Hubble time rapidly shrinks as $A$ increases beyond 8 \rsun
\end{itemize} 

To summarize: (i) the great majority of the systems merging within a Hubble time have initial separations in a small range, (ii) the total mass of the DNS system also varies within a small range, and (iii) for any $(A,\mdn)$ combination, a wide range of eccentricities yields the same value of \taugw. To the aim of describing the general properties of the distribution of the GWR delays, it appears then appropriate to adopt continuous parametrized expressions for the distributions of $(A,\mdn,e)$, to be folded with Eq. (\ref {eq_taugw}). In this way we aim at characterizing  the distribution of \taugw\ and its dependence on the various astrophysical parameters, identifying the most important ones. 

\section{Montecarlo Simulations for the distribution of the GWR delays}

We adopt power law distributions for the separation, binary mass and eccentricity:

\begin{equation}
\begin{cases}
f(A) \propto A^{\beta} \\
f(\mdn) \propto \mdn^{\gamma} \\
f(e) \propto e^{\rho} .
\end{cases}
\label{eq_distr}
\end{equation}

The choice for the distribution of $A$  finds some support from the numerical results of BPS computations: for example in  \citet{giacobbo18} the distributions of the separations of DNS systems which merge within a Hubble time can be described as a power law with exponent $\simeq -1$, and a downturn at separations below $\sim$ 1 \rsun. Also \citet{belczynski18} find a power law distribution for the separations of the DNS systems at formation, albeit with steeper exponent, $\simeq -3$, and no evident downturn at the smallest separations.
The adoption of power law distributions for \mdn\ and $e$ is more arbitrary, but will turn out relatively unimportant for the slope of the DDT.

\begin{figure*}
\centering
\resizebox{\hsize}{!}{
\includegraphics[angle=0,clip=true]{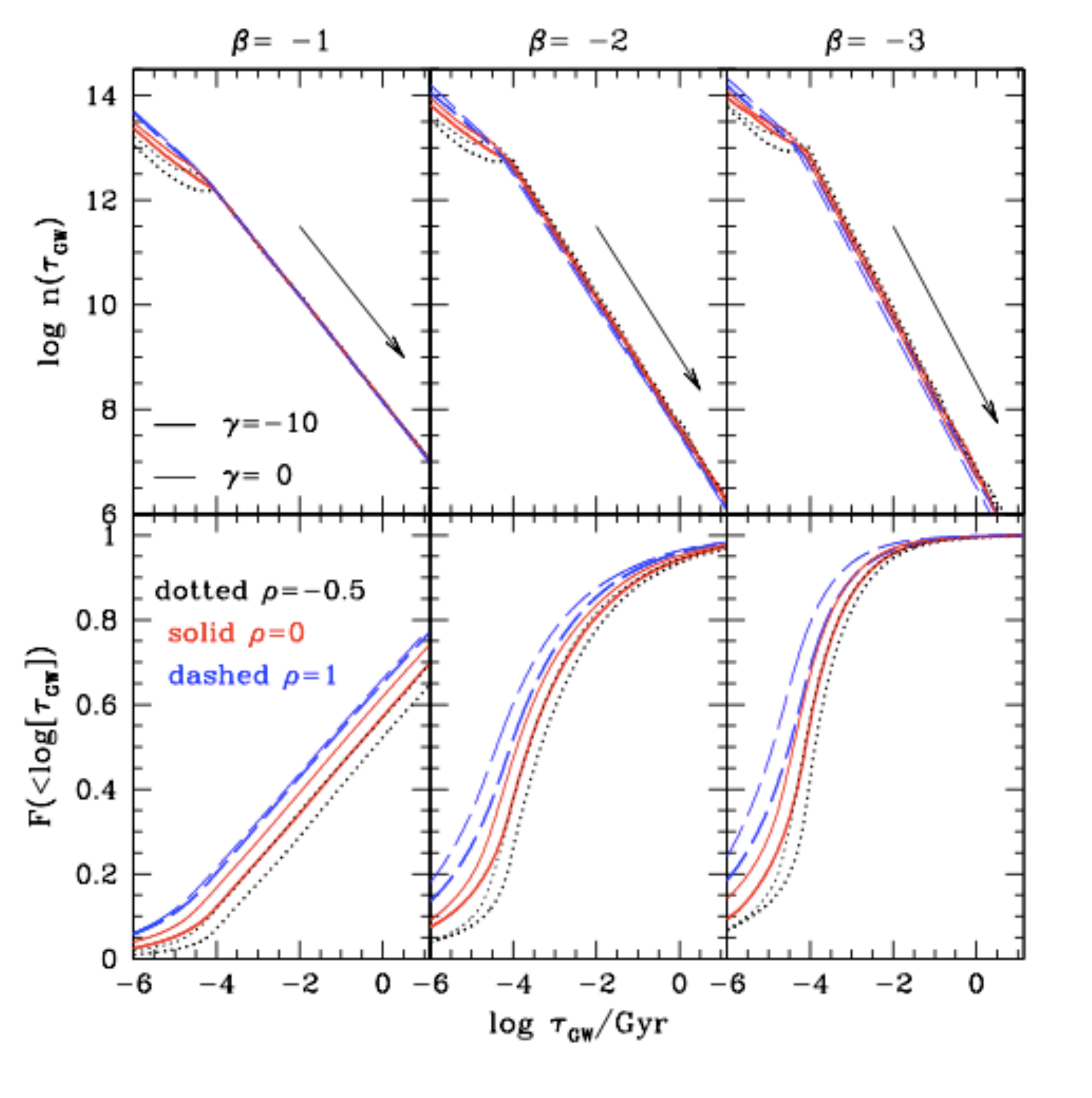}}
\caption{Differential (top) and cumulative (bottom) distributions of the GWR delays from Monte Carlo simulations, assuming that ($A,\mdn,e$) vary independently in $0.2 \leq A/\rsun \leq 30$, $2.2 \leq \mdn/\msun \leq 4$, and $0 \leq e < 1$. The simulations include (6000,500,1000) values of respectively ($A$, \mdn, $e$) for a total of $3\times 10^9$ random extractions. Left, central and right panels show the results obtained with $\beta=-1,-2$ and -3, respectively. The color and line type encode the parameter $\rho$, while the line thickness encodes the parameter $\gamma$, as labelled in the left panels. The arrows in the top panels show power laws with slopes of $s=-1,-1.25,-1.5$ from left to right.}
\label{fig_ntg_unc}
\end{figure*}

\begin{figure*}
\centering
\resizebox{\hsize}{!}{
\includegraphics[angle=0,clip=true]{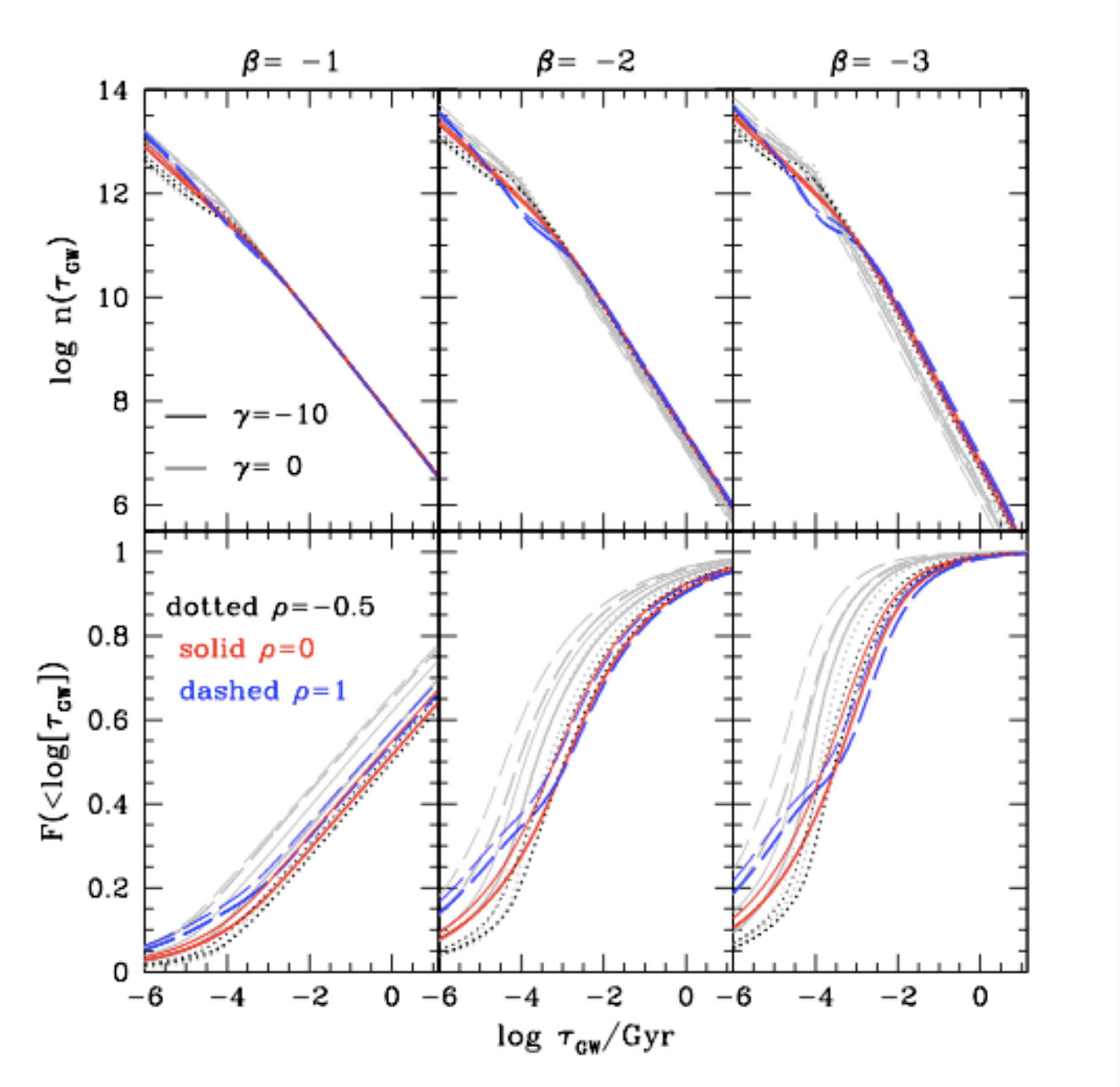}}
\caption{The same as in Fig. \ref{fig_ntg_unc} but for Monte Carlo simulations which incorporate a description on the effect of the supernova kick on the orbital parameters (see text). The simulations include $10^6$ values of  (\ai,$e$) pairs and 1000 values of \mdn, for a total of $10^9$ random extractions. In grey we plot the results of the Monte Carlo simulations shown in Fig. \ref{fig_ntg_unc}. The distributions in the top panels have been scaled to the same total number of extractions of $10^9$.}
\label{fig_ntg_A3}
\end{figure*}

\begin{figure*}
\centering
\resizebox{\hsize}{!}{
\includegraphics[angle=0,clip=true]{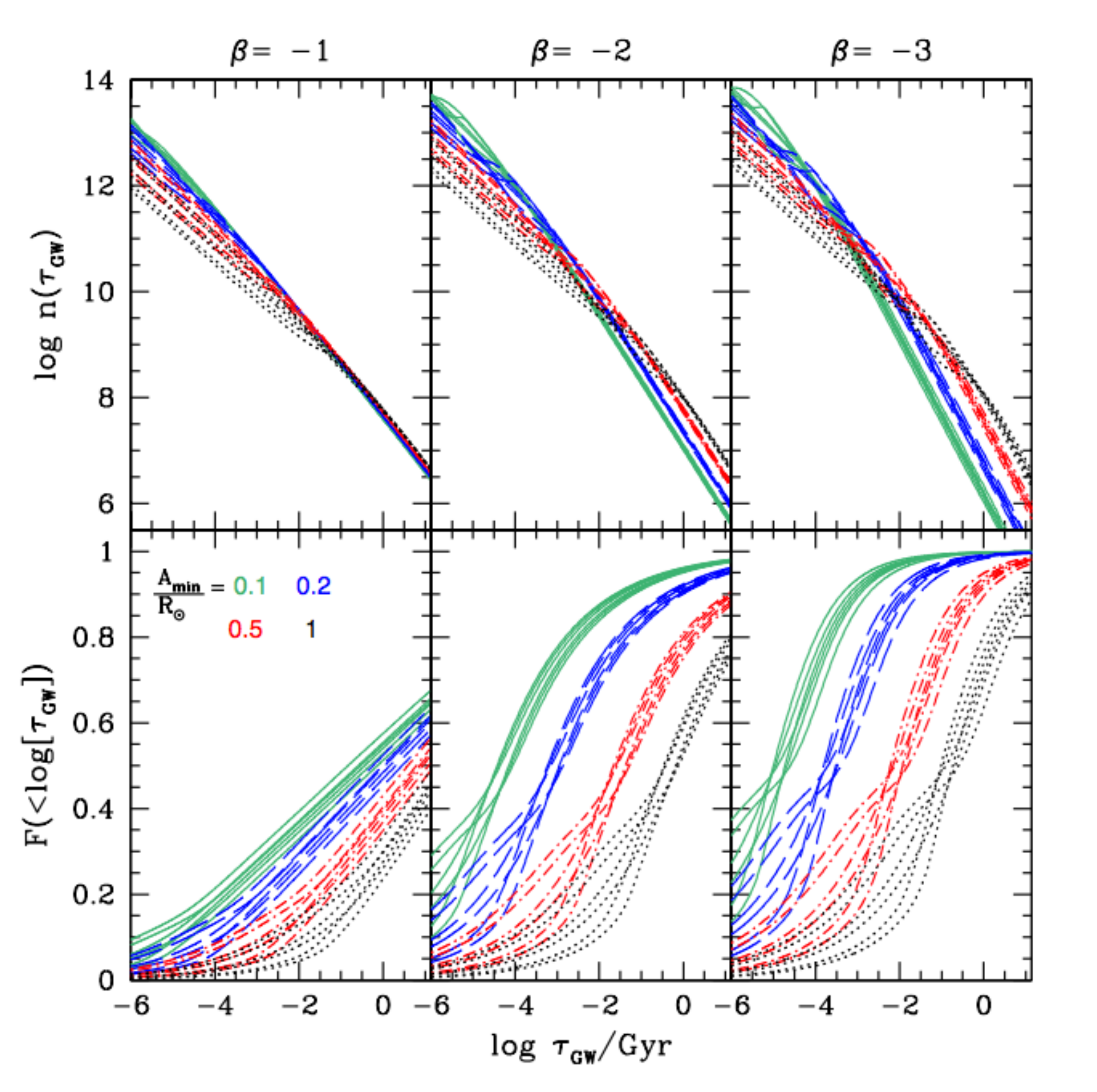}}
\caption{Differential (top) and cumulative (bottom) distributions of the GWR delays from Monte Carlo 
extractions for different choices of the minimum separation of the binary system prior to the second supernova explosion: $\ain/\rsun$ = 0.1,0.2,0.5 and 1 plotted respectively green (solid), blue (dashed), red (dot-dashed) and black (dotted) lines. All simulations adopt the same maximum value for the separation $\aix = 50 \rsun$. Each simulation comprises $10^6$ values of  (\ai,$e$) pairs and 1000 values of \mdn, for a total of $10^9$ random extractions, and includes a description of the effect of the supernova kick.
For each value of \ai\ we plot 6 lines, each corresponding to a choice for the parameters of the distribution of \mdn\ ($\gamma=-10,0$) and of the eccentricity ($\rho=-0.5,0,1$).
The left, central and right panels show the results for different slopes of the distribution of separations \ai\ as labelled on top.}
\label{fig_ntg_Arange}
\end{figure*}

\subsection{Results for independent variables }

Figure \ref{fig_ntg_unc} shows the distribution of the GWR delays under different options for the exponents ($\beta,\gamma,\rho$), in the hypothesis that $A$,
\mdn\ and $e$ are independent variables.
We consider $\beta$ varying between $-1$ and $-3$ on the basis of the results of BPS realizations mentioned above, while the tested values of the parameters $\gamma$ and $\rho$ are meant to explore the general response of the distribution of \taugw\ to the distributions of binary mass and eccentricity. The two cases $\gamma=0,-10$ show the effect of going from a flat distribution of \mdn\ to a case in which the great majority of systems are found at the low mass end; the three values of $\rho$ depict the effect of varying the distribution of eccentricities from a function favouring low ($\rho=-0.5$) to  one favouring high ($\rho=1$) values of $e$. 

At delays longer than 0.1 Myr the distribution of \taugw\ is very well described by a power law with an exponent $s \simeq -1,-1.25$ and $-1.5$ respectively for $\beta=-1,-2$ and $-3$, while the parameters $\gamma$ and $\rho$ have a negligible effect. In the appendix we show that, under some simplifications, the distribution of \taugw\ can be derived analytically, and  results in a power law with exponent  $s=0.25\times \beta -0.75$ modified at short delay times. In fact, the lower limit adopted for the separation $A$ implies a dearth of fast merging systems, causing a flattening of the distribution at short \taugw. The effect is amplified when lower values of $\gamma$, which disfavour massive systems, are assumed. Similarly, the flattening is more pronounced for lower vaues for $\rho$, which imply a larger fraction of low eccentricity systems. 

These patterns appear well visible in the cumulative distributions of the GWR delays (bottom panels). For $\beta=-3$, 90\% of the systems merge within $\sim$ 1 Myr from the formation of the second neutron star, while for $\beta=-1$ only $\sim$ 70 \% of the systems merge within a Hubble time. Therefore, the
timescale for the release of nucleosynthetic products from kilonovae is extremely  sensitive to this parameter. The cumulative distributions also emphasize the dependence of the results on $\rho$ and $\gamma$, with a larger fraction of systems at short \taugw\ obtained with the higher values of $\rho$ and $\gamma$. 

\subsection{Effect of the supernova kick}

As mentioned above, the distributions in Fig. \ref{fig_ntg_unc} have been computed assuming that $A,\mdn$ and $e$ are independent variables. However, in real DNS systems one may expect that the larger separations are coupled to higher eccentricities as a consequence of the supernova kick. The effect on the orbit of the DNS system induced by the kick due to an asymmetric explosion of the second supernova has been studied in detail by \citet{kalogera96} and more recently by \citet{andrews19}. In the latter paper, the authors perform simulations of binaries formed by a neutron star plus a massive Helium star which undergo a kick when the Helium star explodes. The simulations are computed for a distribution of supernova kicks, Helium star masses and separations of the system. The results show that the ratio between the final and the initial separation (\afai) and the eccentricity are not uniformily distributed, but rather cluster around two loci: $\afai = (1 \pm e)^{-1}$. Thus, it appears that the effect of the kick is that of promoting a relation between the separation and eccentricity of DNS systems: some on the branch $\afai=(1+e)^{-1}$ (branch 1) characterized by a general shrinking of a factor up to 2 , others on the branch $\afai=(1-e)^{-1}$ (branch 2), which can turn out very wide, but at the same time with large eccentrities. Since the GWR delay is very sensitive to both the separation and the eccentricity it is important to explore the effect of these relations.
Therefore, we performed another set of Monte Carlo simulations in which we extract pairs of values of ($\ai,e$), to which we associate a value for the separation to be used in Eq. (\ref{eq_taugw}) of $\af=\ai/(1 \pm e)$, the two branches randomly represented in equal proportions. The parameters ($\ai,e$) are considered independent and follow the power law distributions with exponents $\beta$ and $\rho$, respectively. The results of this set of Monte Carlo extractions are shown in Fig.4.
Systems which belong to branch 1 end up with a relatively short \taugw\ due to the smaller separation, while systems which belong to branch 2 have a long \taugw\ because of the larger $\af$. In our experiment the second effect prevails so that the overall distribution becomes more populated at the long GWR timescales compared to what obtained when the variables are assumed independent (see grey lines in Fig. \ref{fig_ntg_A3}). The differential distributions still appear  well represented by a power law with slope $s= 0.25 \beta - 0.75$ at $\taugw \gtrsim $ 1 Myr, but the flattening of the distributions at early times is more pronounced, and more systems merge on longer GWR delays. The cumulative distributions (bottom panel) better illustrate the change, with 50 \% of the systems merging within 1 Gyr if $\beta=-1$. Steeper distributions of \ai\ yield much shorter timescales, but still longer than those obtained for the case of independent variables, for the same value of $\beta$. It also appears that in this set of simulations the relation between $e$ and $\afai$ reduces the sensitivity of the distributions on the $\gamma$ and $\rho$ parameters, the cumulative curves running close to each other, especially at $\taugw \gtrsim 0.1$ Myr.\\

Given the sensitivity of the GWR delay to the separation, the limits on this parameter impact on the resulting distributions. We have then performed more simulations varying the minimum (\ain) and maximum (\aix) separation of the binary systems before the second supernova explosion. In these simulations we adopt the scheme which describes the effect of the supernova kick on the orbit which seems more akin to actual astrophysical situations. Fig. \ref{fig_ntg_Arange} shows the effect of varying \ain, while varying
\aix\ has a weak impact, as we show in the appendix. The differential distributions (top panels) still appear to follow two regimes: a power law with slope $s= 0.25 \beta - 0.75$ at relatively long GWR delays, and a flatter relation at short delays. The value of \taugw\ at which this transition occurs   gets shorter and shorter as \ain\ decreases, because of the higher number of systems with small values of \ai. For the same reason, as \ain\ decreases, the fraction of early merging increases. For $\beta=-1$ the fraction of systems merging within 1 Myr varies from $\sim 0.05$ to $\sim 0.3$ as \ain\ decreases from 1 to 0.1 \rsun. As the distribution of the separations steepens ($\beta$ decreases) more and more systems are born with small separations and their merging timescales become shorter and shorter. For $\beta=-3$ the fraction of systems merging within 1 Myr goes from $\sim$ 15 \% to $\sim$ 90 \% as \ain\ decreases from 1 to 0.1 \rsun. 
Notice however that these figures result from having described the distribution of the separations as a pure power law, which maximizes the number of systems at the low \ai\ values by construction.

\subsection{The distribution of the GWR delays}

The results of the simulations can be summarized as follows:
\begin{itemize}
\item The distribution of the GWR delays is mostly sensitive to the distribution of the separations of the binary systems when the second neutron star is formed. Describing the latter with a power law with exponent $\beta$, the distribution of the GWR delays is also a power law with exponent $s=-0.75+0.25 \times \beta$ for delays longer than some characteristic \taugw\ ($\lesssim$ 10 Myr).
\item The minimum value of the separation is a crucial parameter for the fraction of systems with short merging timescales; its impact depends on how steep the distribution of \ai\ is; in the extreme combination \ain=0.05 \rsun and $\beta=-3$ we find that all DNS systems merge within 1 Myr from their formation.
\item The dependence of the distribution of the GWR delays on the distribution of the other variables, (\mdn\ and $e$) is less pronounced. However, in general, the larger the fraction of massive and/or eccentric binaries, the larger the fraction of systems merging on a short timescale. 
\end{itemize}

Although the distributions of the three variables ($A,\mdn,e$) will not be pure power laws, we believe that our simulations explore a sufficiently wide parameter space to derive crucial characteristics of the distribution of the GWR delays. Concerning the value of the \ain\ parameter, we remark that  
the radius of Helium stars progenitors of a neutron star ranges between $\sim$ 0.3 and 1 \rsun\ \citep{stan19} during the central Helium burning phase. Systems with separations smaller than several tenths of \rsun\ are likely to undergo mass exchange before Helium ignition, thereby avoiding the successive nuclear burnings which lead to the supernova explosion. For systems in which the Helium star component completes core Helium burning inside its Roche Lobe, further orbital shrinking could however occur during the evolution after central Helium exhaustion, when the Helium star expands. Models by \citet{laplace20} show that the final separation of the system could be as low as $\sim$ 0.07 \rsun. Therefore we also computed models adopting \ain = 0.05 \rsun. 
It appears however very unlikely that the distribution of the separations of DNS systems is a power law all the way down to such a small value; nevertheless this extreme case allows us to check the results of chemical evolution models when the nucleosynthetic contribution from kilonovae occurs on an extremely short timescale.

\begin{figure}
\includegraphics[width=\columnwidth]{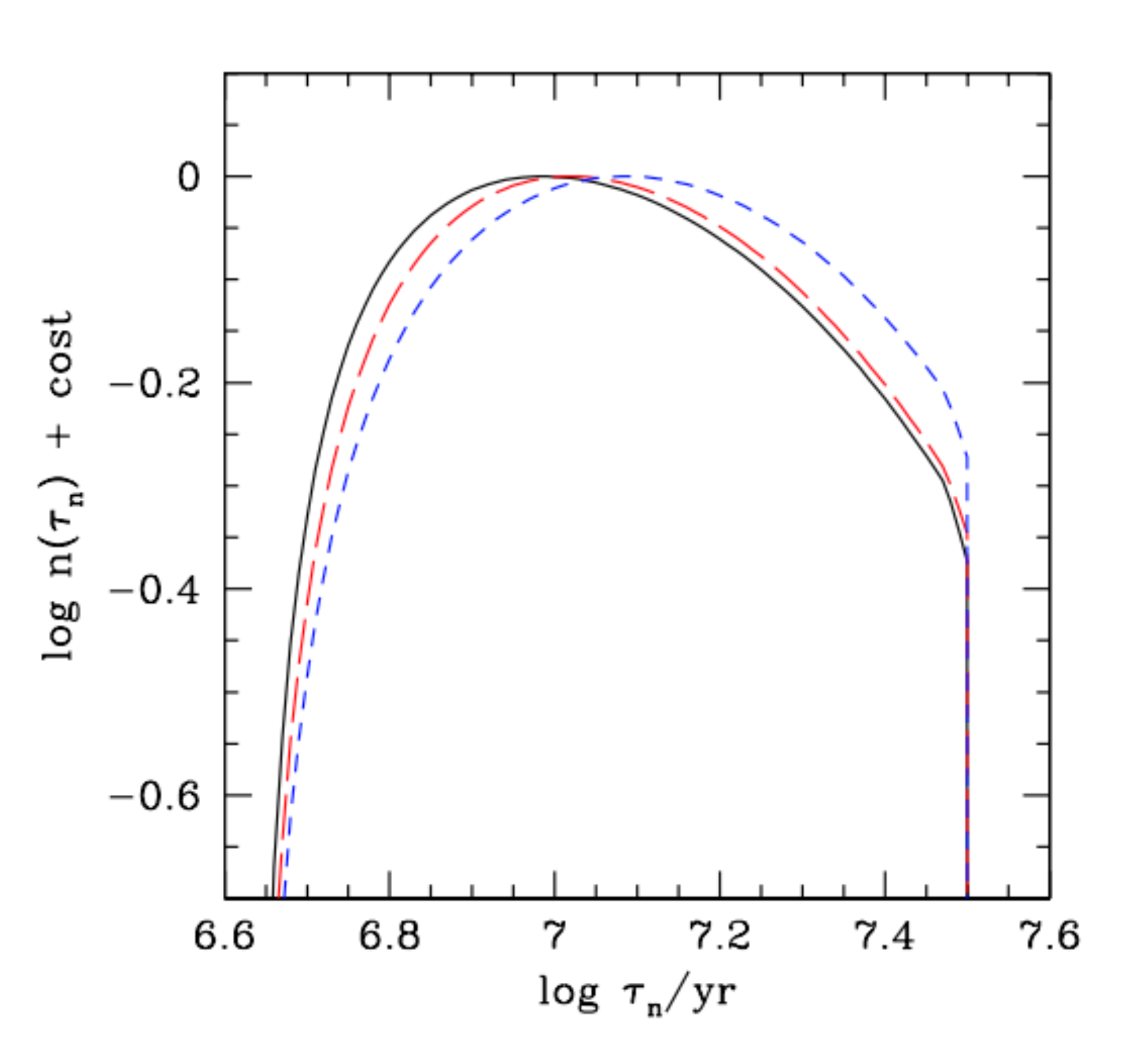}
\caption{The distribution of evolutionary lifetimes of the secondary in DNS system  progenitors for 3 choices of the distribution of the primary masses and mass ratios. Black solid line: $n(\mpri) \propto m^{-2.35}, f(q) \propto q$ (the combination adopted here); red, long dashed line: 
$n(\mpri) \propto m^{-2.3}, f(q) \propto q^{-0.1}$ (the combination in \citet{giacobbo18}); blue, short dashed line:  $n(\mpri) \propto m^{-2.6}, f(q) \propto q^0$ (a steep IMF combined with a flat distribution of the mass ratios).}
\label{fig_ntaun}
\end{figure}

\begin{figure*}
\centering
\resizebox{\hsize}{!}{
\includegraphics[angle=0,clip=true]{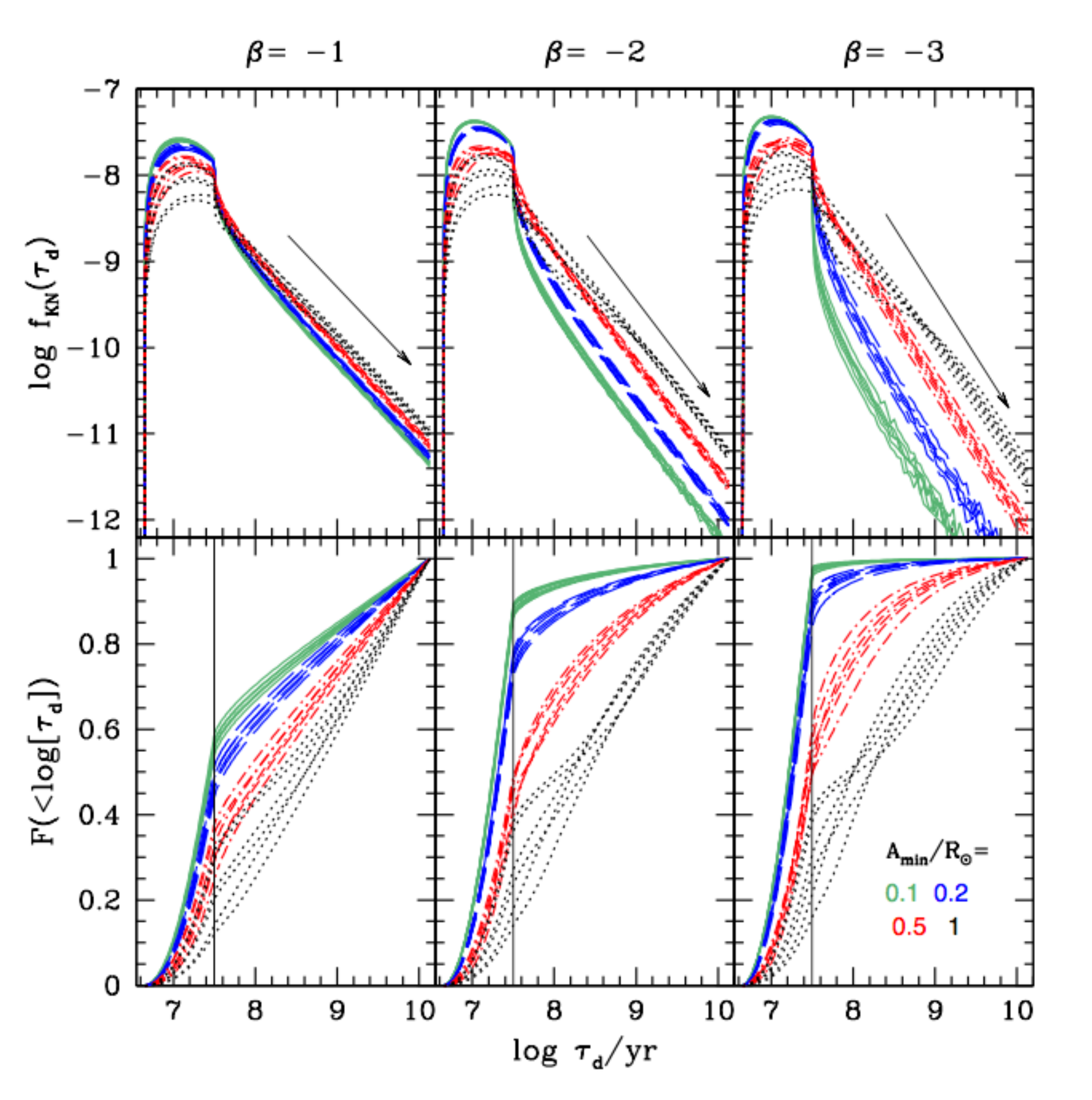}}
\caption{Differential (top) and cumulative (bottom) distributions of the total delay times of kilonova explosions for different distributions of the separations, DNS total masses and eccentricity, in the same fashion as in Fig. \ref{fig_ntg_Arange}. The arrows in the top panels show power laws with slope of $s=-1,-1.25,-1.5$ from left to right. The vertical line in the bottom panels is drawn at a delay time of 32 Myr, i.e. the evolutionary lifetime of the least massive neutron star progenitor adopted here. 
}
\label{fig_ddt}
\end{figure*}

\begin{figure*}
\centering
\resizebox{\hsize}{!}{
\includegraphics[angle=0,clip=true]{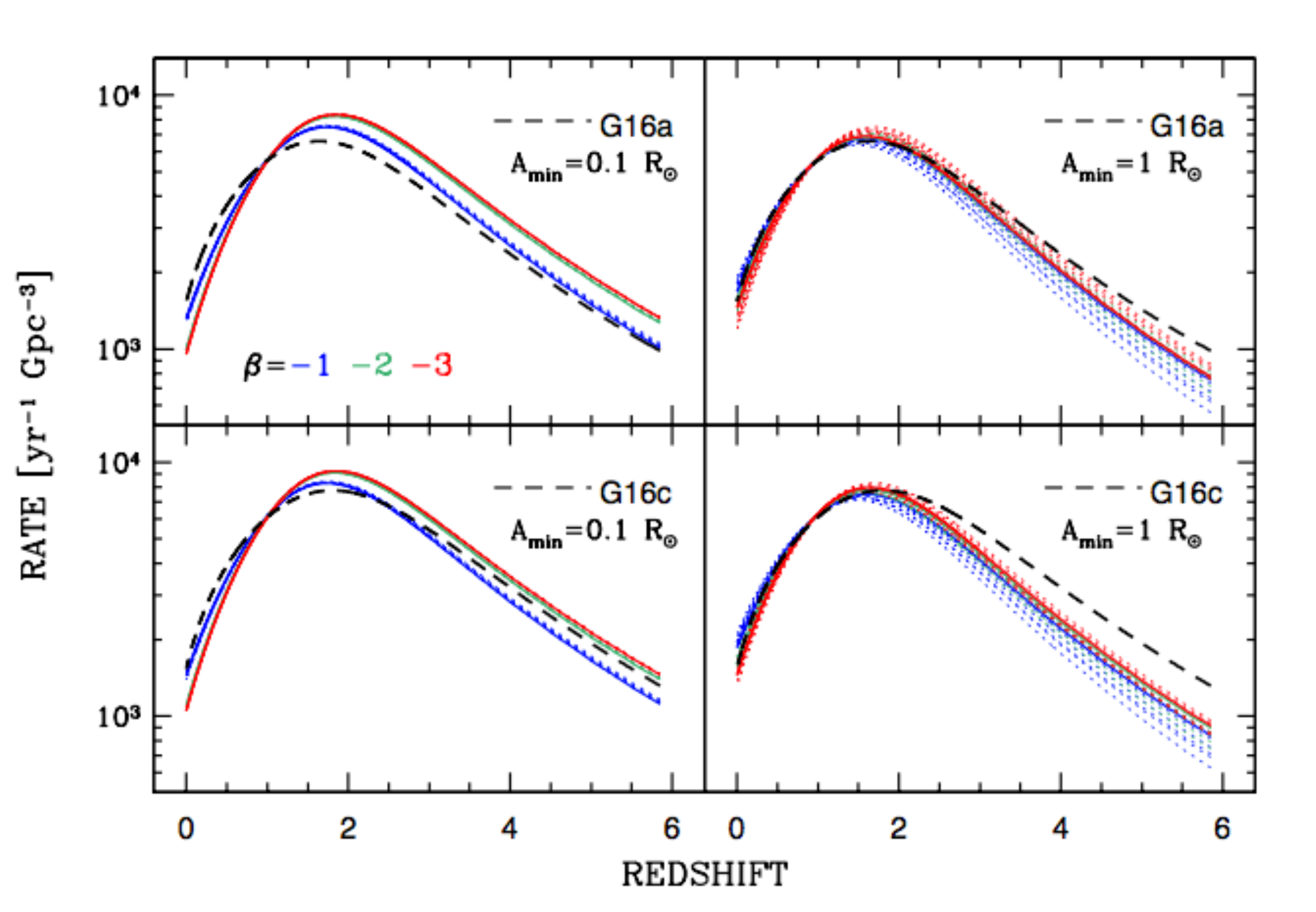}}
\caption{The redshift distribution of SGRBs as in curves $a$ (upper panels) and $c$ (lower panels) in \citet{ghirlanda16} (dashed black lines) compared to the best fitting models obtained convolving the cosmic star formation history with selected DDT distributions (coloured lines). Blue, green and red lines show the models computed with $\beta=-1,-2,-3$ respectively.
Left (right) panels show models which adopt \ain=0.1 (1) \rsun. For each combination of the ($\beta$,\ain) parameters, we plot with dotted lines all the models computed with the different values of $\rho$ and $\gamma$ and highlight the best fitting of them with the solid line. Notice that the dependence of the models on the $\rho$ and $\gamma$ parameters is virtually absent when $\ain=0.1 \rsun$, while for $\ain=1 \rsun$ the distribution of binary masses and eccentricity affect the fitted models. }
\label{fig_rate}
\end{figure*}

\section{The distribution of the total delay time}

As anticipated in Sect. 2, for a DNS system evolving in isolation, the total delay time is the sum of the  evolutionary lifetime of the secondary component with the GWR delay. In this section we derive the differential distribution of the total delay times of merging DNS systems, starting from the cumulative distribution \footnote{We follow this approach because \taun\ and \taugw\ are not independent variables, both being related to the mass of the secondary component of the system.}.

The contribution to the systems with total delay shorter than \taud\ from systems with nuclear delay between \taun\ and \taun$+$d\taun\ is:

\begin{equation}
{\rm d} n(\taud) = n(\taun) \times g(\taun,\taud) \, {\rm d} \taun  
\label{eq_dntaun}
\end{equation}

\noindent
where $n(\taun)$\,d\taun\ is the number of binaries with nuclear delay between \taun\ and \taun+d\taun, and $g$(\taun,\taud) is the fraction of them with GWR delay shorter than ($\taud-\taun$). For each nuclear delay, $g$(\taun,\taud) is the value of the cumulative distribution of \taugw\ presented in the previous section, read off at the appropriate \taugw = $\taud-\taun$.

Summing on the relevant range of \taun\ we derive the total number of systems with delay shorter than \taud\ as:

\begin{equation}
    F(<\taud) = \int_{\tauni}^{\min(\taud,\taunx)}n(\taun)\,g(\taud,\taun) \, {\rm d} \taun
    \label{eq_iddt}
\end{equation}

\noindent
where \tauni\ and \taunx\ bracket the range of nuclear delays which contribute to the kilonova explosions, i.e. the evolutionary lifetimes of their progenitors.

\subsection{The distribution of the nuclear delay}

The distribution of the nuclear delays can be derived as \citep[see,e.g.][]{gr_book}:

\begin{equation}
    n(\taun) \propto n(\msec) \Bigl |\frac{d \msec}{d\taun}\Bigr|
    \label{eq_ntaun}
\end{equation}

\noindent
where \msec\ is the mass of the secondary star with evolutionary lifetime equal to \taun. The distribution function of the secondary masses in the systems of interest can be derived from \citep[see][]{greggio05}:

\begin{equation}
    n(\msec) \propto \int_{\mprin}^{\mprix} n(\mpri) \, f(q) \frac{d\mpri}{\mpri}
    \label{eq_nmsec}
\end{equation}

\noindent
where \mpri\ is the mass of the primary, $n(\mpri)$ is its distribution function, $q$ is the mass ratio (\msec/\mpri), and
\mprin\ and \mprix\ bracket the range of primary masses of interest. Assuming that the mass of neutron star progenitors ranges between 9 and 50 \msun, we have that \mprin/\msun = max(9,\msec) and \mprix/\msun=50.

Fig.\ref{fig_ntaun} shows the distribution function of the nuclear delays computed with Eqs. (\ref{eq_taun}), (\ref{eq_ntaun}) and (\ref{eq_nmsec}) for different distributions of the primary masses and mass ratios (see caption).
The function $n(\taun)$ is defined between a minimum ($\simeq$ 4.5 Myr) and a maximum ($\simeq$ 32 Myr) value of the nuclear delay, respectively the evolutionary lifetimes of the maximum and minimum mass of neutron star progenitors adopted. Within these limits, as \taun\ increases, the distribution first increases, due to the effect of the IMF which provides more and more systems with smaller \msec. This effect is countered by the decrease of the rate of change of the secondary mass (i.e. the $\dot{\msec}$ factor in Eq. (\ref{eq_ntaun})), so that the function shows a wide maximum and then decreases. Note that  the adopted mass limits for the neutron star progenitors are not critical for the shape of the distribution of the nuclear delays. The distribution $n(\taun)$ appears mildly affected by the choice of $n(\mpri)$ and $f(q)$; in the following we adopt a Salpeter IMF for the primary masses and a distribution of the mass ratios which favours the high values of $q$ ($f(q)\propto q$).  

\subsection {The final DDT}

With these ingredients we have computed the distribution of the total delay time of kilonova explosions (\fkn) by first deriving the cumulative distribution from Eq. (\ref{eq_iddt})
and then computing its derivative with respect to the delay time:

\begin{equation}
    \fkn(\taud) = \frac{d}{d\taud}F(<\taud).
    \label{eq_ddt}
\end{equation}

Fig.\ref{fig_ddt} shows the resulting distribution of the delay times, normalized to unity over the range $0 \leq \taud \leq 13.5$ Gyr. All distributions show a strong early peak followed by a decline which can be well described with a power law. The peak is populated by systems merging soon after the formation of the neutron star from the secondary component of the binary.
No merging occurs earlier than \taud $\simeq$ 4.5 Myr, because this is the evolutionary lifetime of the most massive secondary considered here as neutron star progenitor. In the range of delay times between 4.5 and 32 Myr the requirement $\taud=\taun+\taugw$ is met with an increasing range of values of both \taun\ and \taugw. The evolutionary lifetime of the least massive neutron star progenitor considered here is $\simeq$ 32 Myr: mergings at delay times longer than this are achieved only with $\taud-32 \lesssim \taugw/{\rm Myr} \lesssim \taud-4.5$, reflecting the constraint on the nuclear lifetimes of the kilonova progenitors.
The hard limit on \taun\ causes the discontinuity at 32 Myr in the delay times distribution; beyond this limit \fkn\ scales essentially as the distribution of the GWR delays. The same argument was elaborated in \citet{greggio05} for the distribution of the delay times of SNe Ia. We remark that the DDT is proportional to the event rate for a single burst of star formation. Similar to the case of SNe Ia, the rate of DNS merging per unit mass will be large in young and low in old stellar systems, the actual value being the result of the DDT weighted by the star formation history of the system.

Fig. \ref{fig_ddt} shows that the DDT depends mostly on the parameters $\beta$ and \ain\ which characterize the distribution of the separations of the binaries when the second supernova explodes. The peak at short delay times is stronger for distributions which are more populated at low values of \ai\, either  because of a steeper $\beta$, or because of a smaller limit $\ain$, or both. The dependence on the parameters $\gamma$ and $\rho$ which characterize the distribution of binary masses and eccentricities is small, especially so when the distribution of the separations is relatively flat. In the upper panels of Fig. \ref{fig_ddt} we also plot an arrow which represents a power law with exponent $s=-0.75+0.25 \beta$. In most cases the DDT decline follows this power law at delay times longer than $\simeq 0.1$ Gyr, but for some combinations of the parameters this regime sets at a later epoch. Actually, our adopted scheme to describe the effect of the supernova kick impacts on the distribution of the coalescence delays enhancing the fraction of systems with late merging timescales (see Fig.\ref{fig_ntg_A3}). The effect is stronger for lower values of $\beta$ and higher values of $\rho$.

In the lower panels of Fig. \ref{fig_ddt} the vertical line is drawn at 
$\taud=32$ Myr, which can be taken as a partition between \prompt\ and \delayed\ events: \prompt\ kilonovae release their products on a timescale close to that of their progenitors, \delayed\ kilonovae do so after their supernova progenitors have polluted the ISM. The fraction of \prompt\ and \delayed\ events impacts on the chemical evolution of the system by leaving their imprint on the abundance ratios.
The fraction of \prompt\ events is very sensitive to the parameters $\beta$ and \ain, but notice that as $\ain$ decreases this dependence greatly mitigates, and the lower limit to the distribution of $\ai$ becomes unimportant. For $\beta = -3$ almost all systems merge within 32 Myr for $\ai$ smaller than 0.2 \rsun. Conversely, a flat distribution of the separations of DNS system and/or a large lower limit \ain\ imply a late pollution from kilonovae with respect to that of their supernovae predecessors.

\section{The cosmic rate of merging DNS and average efficiency of Kilonovae production}  

In order to compute the evolution of the rate of merging neutron stars in stellar systems, we need to estimate the efficiency of production of these events by a stellar population, or the number of merging neutron stars from a stellar population of unitary mass (\kkn). To evaluate this quantity we consider the cosmic rate of SGRBs , as well as the rate of kilonovae estimated by \citet{abbott20}. 
The rate of merging at epoch $t_0$ in a system experiencing a star formation history $\psi(t)$ is given by:

\begin{equation}
    \Rkn(t_0) = \int_{0}^{t_0} \kkn\ \, \psi(t-\taud)\, \fkn(\taud)\, {\rm d}\taud
    \label{eq_ratekn}
\end{equation}

\noindent
where \fkn\ is the distribution of the delay times of merging double neutron stars, and the integration extends over all the successive stellar generations occurred in the system. The efficiency \kkn\ could well depend on time, e.g. because of IMF variations, and/or because of the metallicity evolution which may impact on the paths of close binary evolution, as well as on the distribution of initial binary parameters. We take a simplified view neglecting the potential variations of \kkn, and proceed evaluating an average value of this efficiency which describes the observed cosmic rates. 

The DDTs presented in the previous section have been normalized to 1 in the range of delay times $0\leq \taud \leq 13.5$ Gyr. Therefore, Eq. (\ref{eq_ratekn}) becomes

\begin{equation}
    \Rkn(13.5) = \kkn \times < \psi >
    \label {eq_rate13}
\end{equation}

\noindent
where the last term is the average star formation rate in the system over the last 13.5 Gyr. If $\psi$ is a mild function of time, the last term can be estimated as the ratio between the stellar mass in the considered system and the Hubble time. Eq. (\ref{eq_rate13}) yields a handy way to estimate the typical efficiency of the evolutionary channel providing merging DNS systems within a Hubble time, which is needed to meet the observed rate of such events. To some extent, this is applicable to the cosmic star formation history, and to stellar populations in late type galaxies, since their star formation rate is a mild function of time.  
The efficiency \kkn\ is normalized to the total (initial) mass of a stellar population, and can be expressed as the product of the number of neutron star progenitors per unit mass (\kalpha) and the fraction of them members of binary systems which merge within a Hubble time (\alphamns): \kkn = \kalpha $\times$ \alphamns. This notation is convenient for the computation of chemical evolution models. 

To evaluate \kkn\ we fit the redshift distribution of SGRBs with models obtained from Eq. (\ref{eq_ratekn}) adopting the cosmic star formation rate by \citet{madau14} 

\begin{equation}
\psi(z)=\frac{0.015(1+z)^{2.7}}{1+((1+z)/2.9)^{5.6}}\ \ \   \msun \text{Mpc}^{-3} \text{yr}^{-1}
\label{eq_madau}
\end{equation}

\noindent
and our DDTs described in the previous section. The relation between redshift and look-back time adopted is that of the $\Lambda \text{CDM}$ cosmological model using the parameters found by \citet{bennett14} (i.e. $H_0=69.6\,\text{km}\,\text{s}^{-1}\,\text{Mpc}^{-1}$, $\Omega_M=0.286$ and $\Omega_\Lambda=0.714$).
As observational constraint we select the two curves (models $a$ and $c$) proposed by \citet{ghirlanda16}. These authors  choose the following functional form for the redshift distribution of all events: 

\begin{equation}
\psi_{\rm SGRB}(z)=\psi_{\rm SGRB}(z=0) \times \frac{1+p_1z}{1+(z/z_p)^{p_2}}
\label{eq_g16}
\end{equation}

\noindent
and find the parameters which best fit a variety of observational data, including the observed redshift distribution and energetic properties of the SGRBs. These parameters turn out of ($p_1,p_2,z_p)$ = (2.8,3.5,2.3) and (3.1,3.6,2.5) respectively for models $a$ and $c$.
Other options for the redshift distribution of SGRBs can be found in the literature \citep[e.g.][]{zhang18}, but require  a DDT poor of  prompt events, which is difficult to reconcile with the fact that neutron stars are produced by massive stars which have very short evolutionary lifetimes
\citep[see, e.g.][]{simonetti19}. 
We obtain the redshift distribution in natural units (events per yr per {\rm Gpc$^3$}) by adopting a value of $\psi_{\rm SGRB}(z=0)$ equal to the local rate of merging neutron stars estimated by \citet{abbott20}. Notice that this value is compatible with the estimates of the local rate of SGRBs found in the literature \citep{coward12,petrillo13,fong15} within the (large) uncertainties. To find the best fit value of the \kkn\ parameter we minimize the distance between the constraining curves and the models in $0 \leq z \leq 2$, which is the redshift range covered by events in the \citep{ghirlanda16} sample.

Fig. \ref{fig_rate} shows the results of our fitting procedure using DDTs with the three values of $\beta$ and two extreme values for the minimum separation of the DNS systems at birth. The combination ($\beta=-3,\ain=0.1$), shown in red on the left panels of Fig. \ref{fig_rate}, favours prompt mergings; the combination ($\beta=-1,\ain=1$), shown in blue on the right panels of Fig. \ref{fig_rate}, favours mergings at late epochs. In spite of the very different values of the astrophysical parameters characterizing the chosen DDTs, all plotted models yield an acceptable representation of the empirical curves, considering that the latter are not directly measured rates, but rather the result of a fitting procedure applied to observational data, which brings along some uncertainty beyond what indicated with the two solutions $a$ and $c$. Models with a small value of \ain\ are characterized by a relatively steep rise of the rate as the redshift increases from $z=0$ to the peak at $z=2$, followed by a milder decrease towards higher redshift, while models with $\ain=1 \rsun$ show the opposite trend. Since the empirical redshift distribution is not well constrained at redshifts larger than the peak, it is not possible to draw conclusions from this comparison. It seems however that models with $\beta=-1$ (in blue) better describe the empirical curves in all cases, except for the case plotted in the lower right panel, where curve $c$ appears to require a steeper distribution of the separations to compensate for the relatively large $\ain (=1\rsun)$. Formally, the minimum distance between models and empirical curves $a$ and $c$, in the range $0\leq z \leq 2$, is obtained respectively for the combinations $(\ain/\rsun,\beta,\rho,\gamma)=(1,-1,1,0)$ and $(0.1,-1,-0.5,0)$. 

From the redshift distribution of the SGRBs we derive an indication in favour of a DDT with a sizable component at long delay times, but the constraining power of this kind of comparison is very limited, because of the shape of the cosmic star formation rate which accomodates stellar populations with a wide range of ages. On the other hand, this property allows us to derive an estimate for the efficiency \kkn\ which is virtually insensitive to the DDT. We find $\kkn =  (6.5 \pm 0.4) \cdot 10^{-5} \msun^{-1}$ for all the models fitted to the \citep{ghirlanda16} curves $a$. Fitting the redshift distribution to curve $c$ leads to a very close value of $\kkn = (7.3 \pm 0.4) \cdot 10^{-5}$.  We acknowledge that this values rest on observational determinations which are affected by a large uncertainty, and that the factor of $\sim$ 10 uncertainty on the local rate of kilonovae \citep{abbott20} implies the same uncertainty of \kkn. We also acknowledge that the approximations introduced to determine \kkn\ weaken its realiability; nevertheless we regard our result as robust, since it is derived under a wide variety of possibilities for the DDT. 
The cosmic star formation history adopted for the fit assumes a Salpeter IMF, with $\kalpha=0.006 \msun^{-1}$ stars with mass between 9 and 50 \msun; therefore the measured  cosmic rate of merging neutron stars requires that, in  a single stellar population, $\simeq 1 \%$ of neutron star progenitors should be found in binary systems which merge within a Hubble time.

\begin{figure*}
\centering
\subfloat[]{\includegraphics[width=0.5\textwidth]{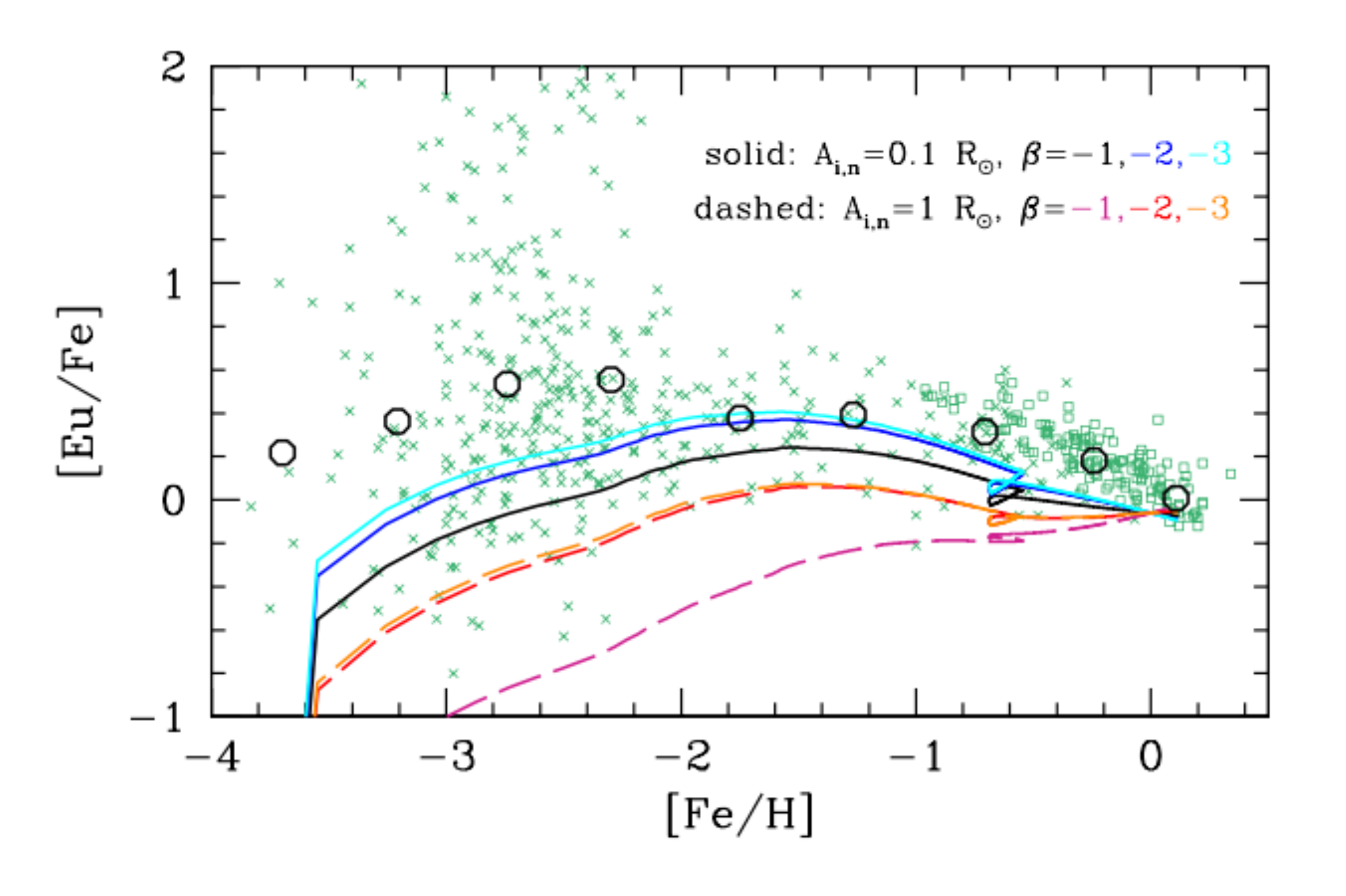}}
\subfloat[]{\includegraphics[width=0.5\textwidth]{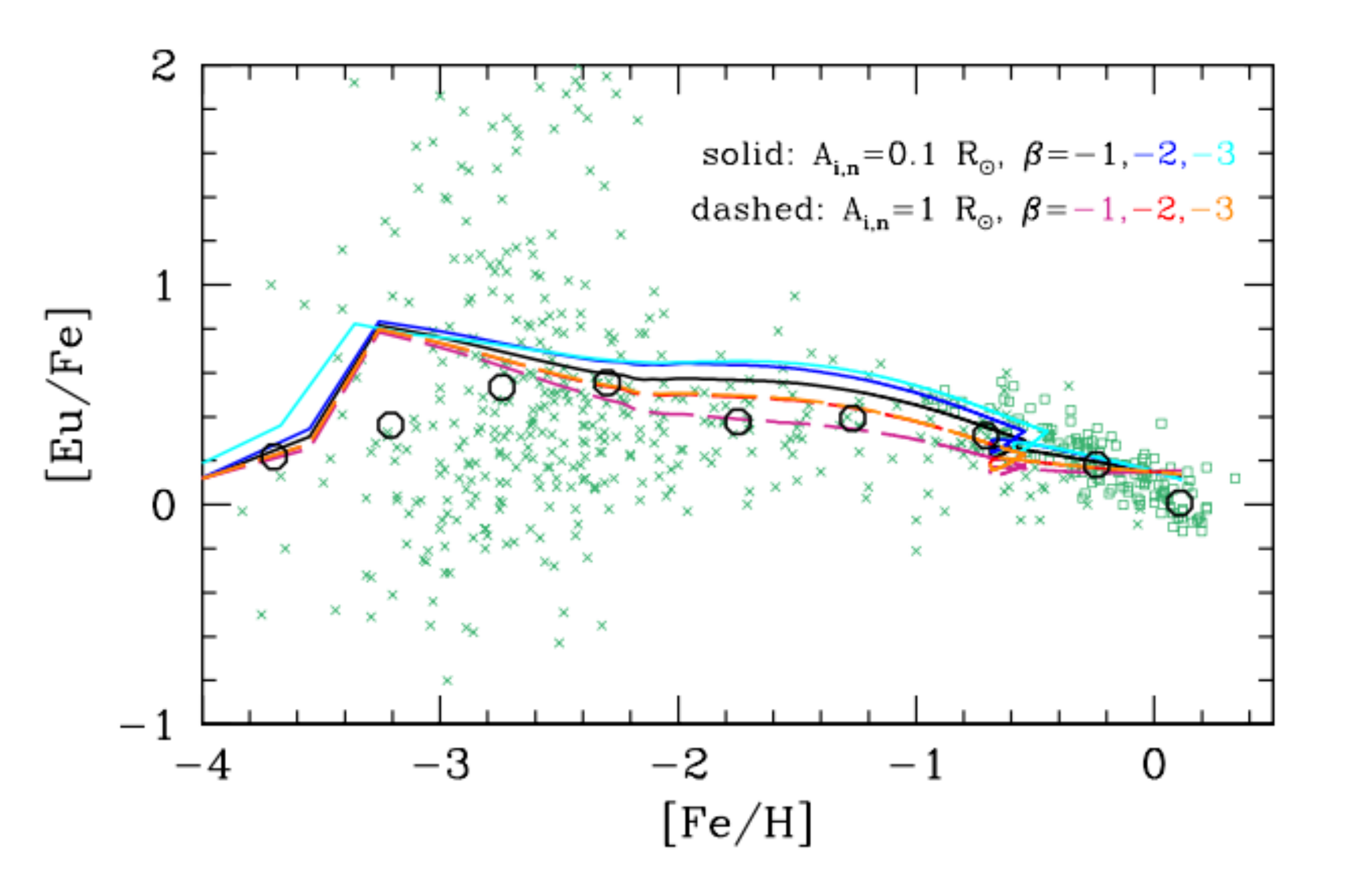}}
\caption{Abundances of Milky Way stars compared to chemical evolution models (coloured lines) computed adopting different distribution of the separations of the DNS systems. Specifically, black, blue and cyan solid lines are obtained with \ain= 0.1 \rsun\ and $\beta= -1,-2,-3$ respectively. In the same sequence, violet, red and orange dashed lines show the results for \ain=1 \rsun. Models in the left panel have been computed adopting kilonovae as the sole source of Europium, while models in the right panel adopt a contribution from CC SNe (see text). Observational data are from   
a compilation of 426 Milky Way halo stars (green crosses) taken from JINABase \citep{abohalima18} and 374 Milky Way thin disk stars (green squares) from \citet{battistini16}. Black circles represent the average values, binned in 0.5 dex wide bins.} 

\label{fig_eufe}
\end{figure*}

\section{The chemical evolution of the Milky Way}

We now consider the constraints on the DDT of merging DNS that can be derived from the chemical properties of Milky Way stars. 
As mentioned in the Introduction, this can be achieved by comparing models for the trend of the [Eu/Fe] abundance ratio with increasing [Fe/H] to corresponding observations. Europium is a pure r-process element and is produced via explosive nucleosynthesis in neutron rich environments. This makes merging neutron stars a very likely site for Eu production. Some Eu may also be synthesized in CC SNe \citep{argast04}, and in rarer, more exotic explosions of very massive stars, i.e. collapsars \citep{winteler12,siegel19}.

\subsection{The model}

The chemical evolution model employed in this work is the classical two-infall model of \citet{chiappini97}, described in detail in e.g.~\citet{matteucci12}, which assumes that the Galaxy formed out of two main gas infall episodes, one giving rise to the halo plus thick disk and the other to the thin disk. The computations follow the time evolution of the gas abundances of 31 chemical elements (from H to Eu), by solving the equations:

\begin{equation}
\label{eq_chem}
\begin{split}
\frac{dG_i(t)}{dt} = & -\Psi(t)\, X_i(t) + R_i(t) + X_{i,0}\, A(t),
\end{split}
\end{equation}

\noindent
written for all the $i$ chemical species. For each element, $G_i$ is the surface density in the gas, normalized to the final total surface mass density (gas plus stars), $X_i$ is its abundance in the gas, and $R_i$ is the rate at which the element is given back to the interstellar medium by winds or stellar explosions.
The first term , i.e. the product between the star formation rate $\Psi(t)$ and $X_i$, describes the rate at which the element $i$ is subtracted from the gas due to star formation, while the last term describes the infall of gas with abundance $X_{i,0}$ at a rate $A(t)$. We adopt the star formation rate by \citet{kennicutt98}, the IMF by \citet{kroupa93}
and a double-exponential infall law  of the two-infall model.  We assume that the first episode of gas accretion occurs on a timescale of the order of 1 Gyr whereas the second on a timescale of 7 Gyr. The abundances of the infalling gas, $X_i(t,0)$, in both cases are assumed to be primordial (no metals). The two infall episodes are separated by a gap in the star formation due to the assumption of a gas threshold density for star formation \citep[see][for details] {chiappini97, matteucci12}. This gap creates a little bump in the model evolutionary curves of the chemical abundances, as we will see in the following.	
In our computations the prescriptions for the nucleosynthesis are taken from \citet{karakas10} for low and intermediate mass stars ($0.8 \le M/M_{\odot} \le 8$), from \citet{doherty14a, doherty14b} for super-AGB stars ($8-9 M_{\odot}$), from \citet{nomoto13} for CC-SNe  ($M \ge 10 M_{\odot}$).
SNe Ia are important contributors of iron and thus have a strong impact on the chemical evolution of galaxies. In our model, for the SNe Ia we use the DDT derived by \citet{greggio05} in the wide double-degenerate scenario, with $\beta=-0.9$, in combination with a production efficiency of $k_{\rm Ia} = 2.5 \times 10^{-3} \msun^{-1}$, fixed by the requirement of reproducing the current rate in the Milky Way estimated by \citet{li11}.
This value is larger by a factor of $\sim 2.5$ with respect to that derived in \citet{greggio19}, partly because of the different IMF and law for the star formation history adopted, partly because of the different method employed to evaluate it. 
\citet{greggio19} consider the correlation between the SNe Ia rate and the color of the parent galaxy from various SN surveys to derive the value of $k_{\rm Ia}$ which best reproduces the observed level of the SNe Ia rate in galaxies of intermediate colors. Therefore, their value of $k_{\rm Ia}$ represents an average realization probability in galaxies. The value adopted here, instead, has been obtained specifically for the Milky Way, under the adopted prescriptions for its star formation history. The chemical yields from SNe Ia are from \citet{iwamoto99}, and from \citet{jose98} for nova systems. Concerning the nova rate, we assume that it is a fraction of the rate of formation of white dwarfs, as first computed by \citet{dantona91} where details can be found.
These prescriptions have been validated through the comparison with a variety of observational data, such as abundance ratios versus metallicity for many chemical elements \citep[D, He, $^{7}\text{Li}$, C, N, O, $\alpha$-elements, Fe-peak elements and heavier, see][]{grisoni18,grisoni19,romano10,romano03}

For the kilonovae explosions we test various possibilities, selecting among each family of model DDTs characterized by a combination of the ($\beta$,\ain) parameters, the one which provides the best fit to the \citet{ghirlanda16} curves, along with its particular value of \kkn. 
The yield of Europium from merging kilonovae is then set to reproduce the solar abundance of this element at the time corresponding to the formation of the Sun, namely 9 Gyr after the starting point of the model, as reported by \citet{lodders09}. We found that this constraint requires a yield of  $5 \times 10^{-6}\, \msun$ of Europium per event. This value is in agreement with both theoretical calculations \citep{korob12} and observations of the AT2017gfo kilonova \citep{tanvir17,troja17}, a fact that supports our determination of \alphamns .
Although the Europium yield from merging neutron stars is uncertain the recent event GW170817 has allowed us to restrict its value to a range of $(2-10) \times 10^{-6}$ \msun.
We have also computed models in which CC-SNe provide a sizable contribution to Europium. In this second case, the yield per DNS merging event has been reduced to $2 \times 10^{-6}\, \msun$, while the yield from CC-SNe has been taken from the SN2050 model of \citet{argast04}. 

\subsection{Results}

Figure \ref{fig_eufe} shows a selection of our models compared to the data. The latter come from 
two databases, one for the  halo stars compiled by \citet{abohalima18} which includes 428 objects (crosses), and another for the generally younger disk stars compiled by \citet{battistini16} with 374 objects (squares). Abundances are expressed using the square bracket notation, where $[\text{X/Y}]=0$ represents the logarithmic ratio between elements X and Y in the Sun. 
The black circles show the average trend obtained by binning the data for individual stars.  The low metallicity stars are characterized  by an average overabundance of [Eu/Fe] $\sim 0.4$ dex;
starting at [Fe/H] $\sim -0.8$, the abundance ratio starts decreasing towards the solar value.
This trend is typical for elements whose production time-scale is shorter than the production time-scale of iron, which in turn is determined by the explosion time of SNe Ia \citep[see e.g.][]{matteucci12}. We notice 
the large spread of the  [Eu/Fe] ratio in the low metallicity halo star. This is likely due to the inhomogeneous pollution of the interstellar medium in the early stages of the star formation history of the Milky Way, as investigated by e.g. \citet{cescutti15} and \citet{wehmeyer15}, an effect which is not captured by our homogeneous model.

The models plotted on Fig. \ref{fig_eufe} encompass the range of possibilities for the DDT of merging DNS envisaged in our approach: for each value of the parameter $\beta$, we show models with the two extreme values of the minimum separation of DNS systems (\ain), and among the models with the same ($\beta$,\ain) parameters we plot the one with the minimum distance from curve $a$ of \citet{ghirlanda16}. We remark that models best fitting curve $c$ by \citet{ghirlanda16} present a very similar trend on this plot. The small loop present in all the lines at [Fe/H]$\simeq-0.6$ is the result of the hiatus in the star formation activity mentioned in section 6.1. 
Panel (a) refers to the scenario in which merging DNS are the only contributors to the Europium production in the Galaxy, while panel (b) shows the effect of adding a contribution from CC-SNe as detailed before. 
 
 When assuming that merging DNS are the only source of Europium, the data can be roughly reproduced with a very low value of \ain\ in combination with  a steep distribution of the delay times ($\beta \lesssim -2)$. Larger values of \ain\  result in a late contribution of Europium to the ISM so that low metallicity stars are formed out of gas with a low Eu abundance. The effect is amplified for flatter DDTs (e.g. $\beta=-1$). Actually, even for the steepest DDT, the [Eu/Fe] ratio of the models is systematically smaller than the average value of the data, with a larger discrepancy in the low metallicity regime ([Fe/H] $\lesssim -2$). In other words, the data seem to require an early source of Europium in the chemical evolution of the Galaxy.
\citet{reichert20} find a similar indication from the analysis of the chemical pattern in Dwarf Spheroidal galaxies.
 Notice that in our model, the abundance [Fe/H]$=-2$ is reached very early, at $\sim 25$ Myr after the start of star formation. In this early stage the Fe enrichment in the ISM is due to CC-SNe, which explode on a shorter timescale with respect to kilonovae, due to the distribution of the coalescence delays. 
 Even for the steepest DDT ($\ain=0.1 \rsun, \beta=-3$) 30 \% of the explosions of each stellar generation occur with a delay longer than 25  Myr. 
 
  When assuming that also CC-SNe contribute to the Europium pollution of the ISM (panel b) all models nicely reproduce the average trend of the Milky Way stars, including the DDTs with the longest timescale for the kilonovae explosions (\ain=1 \rsun, $\beta =-1$). 
 We notice that the models exhibit a somewhat too shallow trend of the [Eu/Fe] ratio as [Fe/H] increases from $\sim -1$ to 0, with respect to the data, especially in the case of \ain = 1, i.e. for DDTs with long pollution timescales.
 The flattening of the [Eu/Fe] ratio in the disc is, at least partly, the result of adopting Fe yields from CC-SNe of \citet{nomoto13}, which include the so-called hypernovae and produce more Fe than the models by \citet{woosley94} adopted in \citet{matteucci14}. 
 Other possibilities to produce a steep [Eu/Fe] trend in disk stars are discussed in \citet{hoto18}, \citet{cote19} and \citet{shonrich19}. 

\begin{figure}
\includegraphics[width=\columnwidth]{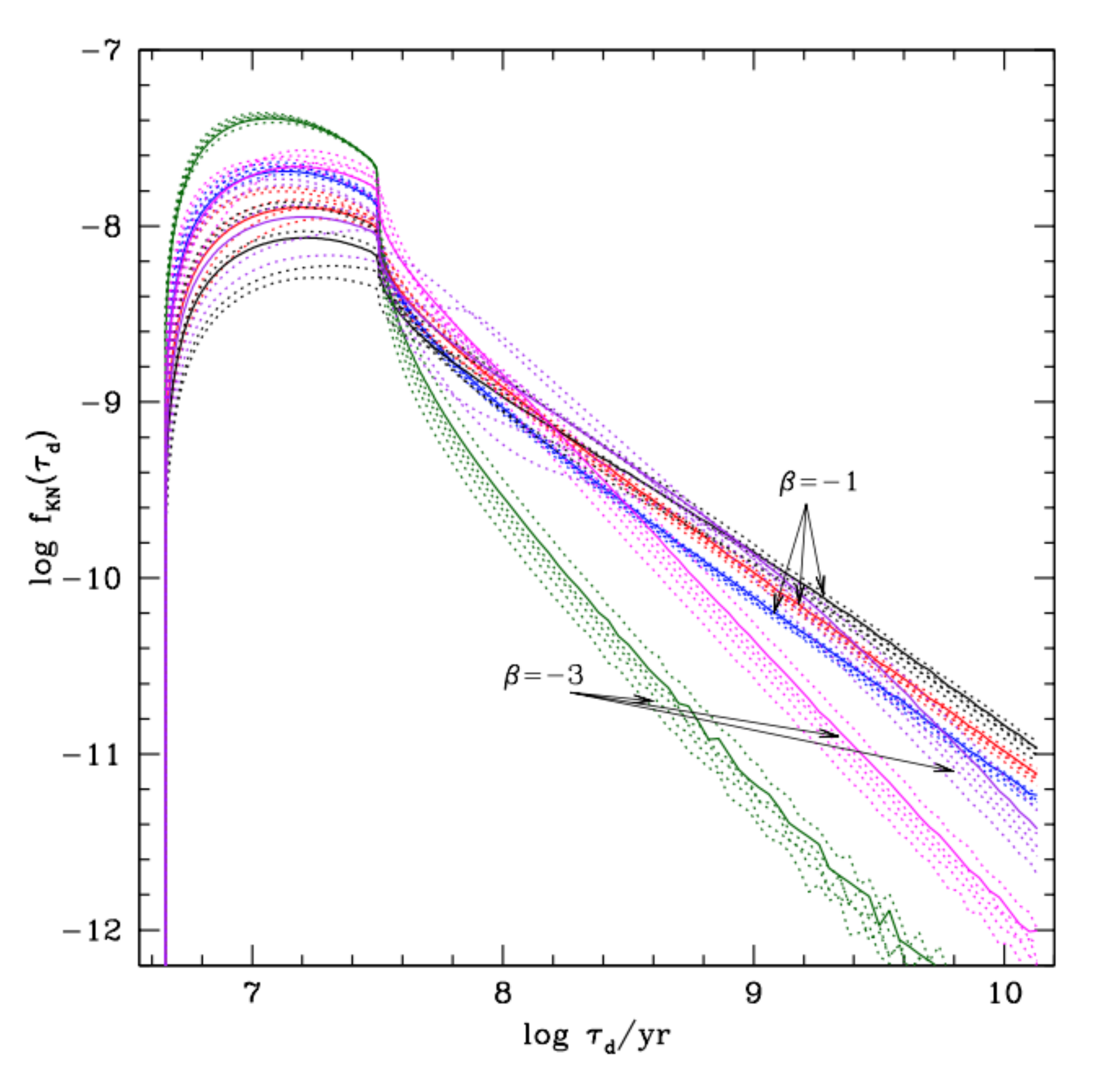}
   \caption{Distribution of the delay times of merging double neutron star systems for a selection of the parameters characterizing the distribution of the separations \ai. For each value of the $\beta$ parameter (indicated with the arrows) we show the distributions obtained with 3 values of the minimum separation \ain= 0.2 \rsun\ (blue and green), 0.5 \rsun\ (red and magenta) and 1 \rsun\ ( black and purple). For each of the $\beta$ and \ain\ options, we plot all combinations of the other parameters, in solid lines we highlight the  ($\gamma,\rho$) = (-10,0) one.}
    \label{fig_summa}
\end{figure}

\begin{figure}
\includegraphics[width=\columnwidth]{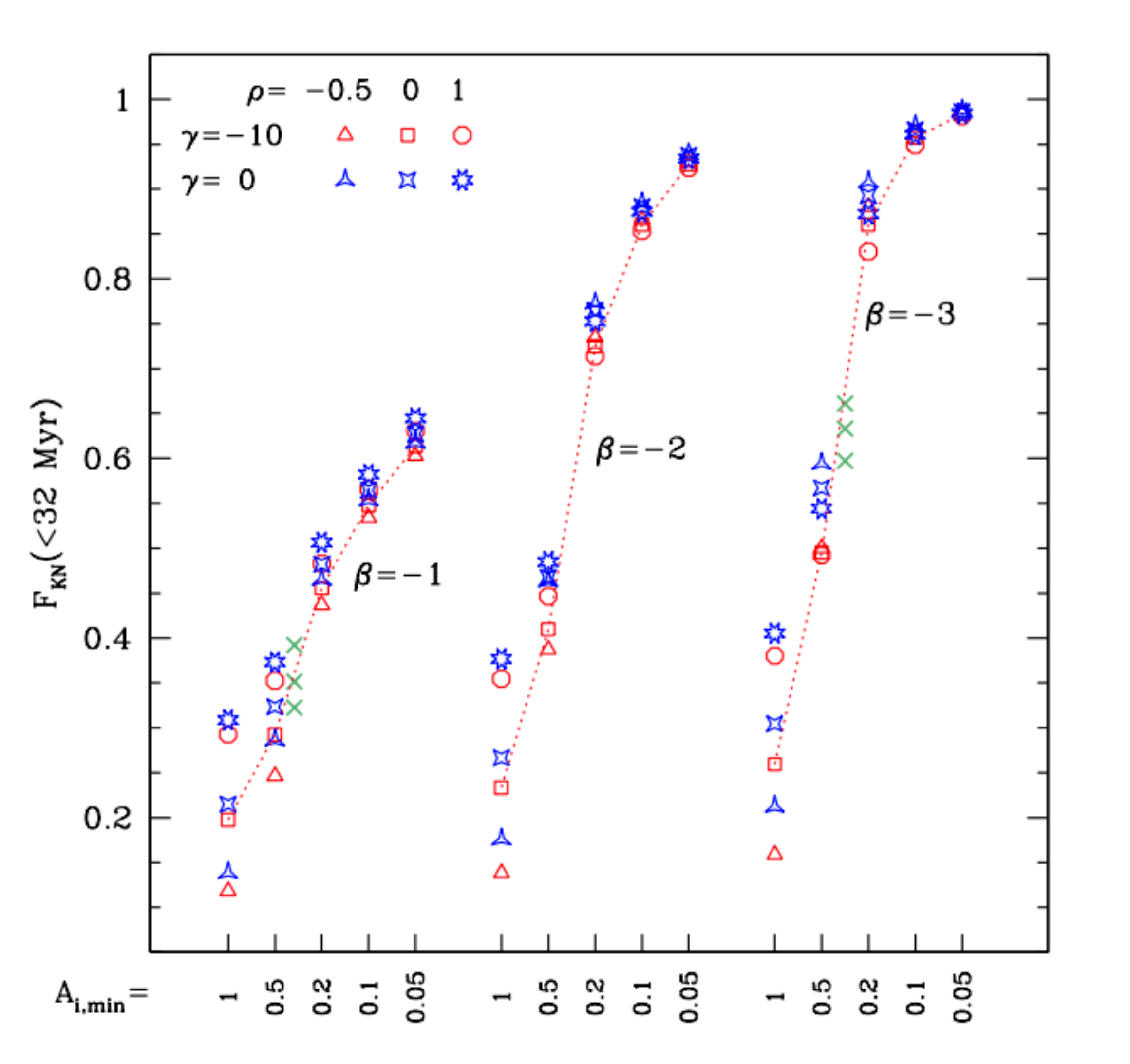}
\caption{Fraction of systems with delay times shorter than 32 Myr for the various cases considered here. For each value of $\beta$ we plot the fractions as function of \ain, reported in the x-axis. The point type encodes the values of $\gamma$ and $\rho$ as labelled. For each $\beta$ the dotted line connects the fractions \Fp\ for the case ($\gamma=-10,\rho=1$) for illustration. Green crosses show the prompt fractions for \ain=0.5 \rsun, in the case of correlation between \mdn\ and \ai. See text for details.}
\label{fig_prompt}
\end{figure}

\section{Summary and Conclusions}

The DDT of merging DNS can be constrained by considering its impact on a number of astrophysical measurements: (i) the cosmic rate of SGRBs, (ii) the chemical abundance pattern of elements synthesized in kilonova explosions, (iii) the relation between the merging neutron stars events and the properties of the parent galaxies. This issue has been considered in various papers \citep[e.g.][]{cote19,simonetti19} to the general conclusion that (i) and (iii) require a sizable fraction of events at late times, while (ii) points to a large number of prompt events, if elements like Eu should be mostly produced by kilonovae. In this paper we further test whether the three observational constraints can be met with a unique distribution of the delay times by considering a variety of astrophysically motivated possibilities for the DDTs. 

In order to figure out these possibilities we focus on the 
 characteristics brought about by the clock which controls the merging event. Since the delay time is the sum of the evolutionary lifetime of the secondary component of the binary system and the coalescence lifetime of the double neutron star, the distribution of the total delay time is a function of the mass of the secondary component and of the total mass, separation and eccentricity of the DNS system.
 We have computed Monte Carlo simulations to derive the distribution of the coalescence timescales assuming that the distributions of the separations, total masses and eccentricity follow power laws with exponents $(\beta,\gamma,\rho)$, respectively. Since the large majority of explosions within a Hubble time are provided by systems with separations smaller than $\sim 8 \rsun$, and
 the total mass of the DNS system ranges between $\sim 2$ and 4 \msun, the assumption of smooth distributions for these variables (in the relevant ranges) seems adequate.
   In the simulations we have implemented a scheme to describe the effect of the kick due to an asymmetric explosion of the second supernova. The distribution of the delays due to the evolutionary lifetimes has been derived analytically, and folding the two distributions we have obtained the distributions of the total delay times for a variety of parameters, which is meant to cover a wide range of astrophysically plausible situations. 

Figure \ref{fig_summa} illustrates the results.

\begin{enumerate}
    \item The distribution of the delay times shows an initial peak in the range 
    $4.5 \lesssim \taud/{\rm Myr} \lesssim 32$ followed by  power law decline. This characteristic stems from the clock controlling the events, which is the composition of the evolutionary lifetime of the secondary star and the time taken by the gravitational waves radiation to bring the system into contact.
     \item The power law exponent which describes the distribution at (relatively) long delay times is close to $s=0.25 \times \beta -0.75$. In most of our explored cases the delay time at which this regime sets in is short ($\sim 0.1$ Gyr), but for some specific combinations of the parameters (e.g. $\beta=-3$, $\ain=1 \rsun$ and $\rho=1$) this epoch becomes as late as a few Gyrs.
    \item The strength of the peak is very sensitive to the parameters $\beta$ and \ain, taken to describe the slope of the distribution of the separations and the minimum separation of the DNS systems when they form.
    \item The parameters $\gamma$ and $\rho$ have a negligible effect on the slope of the DDT, but they affect the strength of the peak. In general, mass distributions which favour high values of \mdn\ and eccentricity distributions which favour high values of $e$ yield more systems with short delay times.
    However, there is a considerable interplay between the three parameters which determine the modifications of the distribution of the delay times  as response to variations of $\beta$, $\gamma$ and $\rho$.
    \item Decreasing \aix\ implies a reduction of the number of mergings at very late epochs, but this effect can be appreciated when \aix\ becomes smaller than a few  \rsun, which seems very unlikely.
\end{enumerate}

\textbf{In the literature it is often assumed that the DDT of merging neutron stars can be described as a pure power law: this is not correct, since at short delay times the DDT is characterized by a plateaux} (item (i) above). 
The width of the plateaux is equal to the difference between the evolutionary lifetimes of the least massive and of the most massive neutron star progenitors. In our computations we have adopted a mass range of $9 \leq M/\msun \leq 50$, which corresponds to a width of $\sim 27$ Myr. 

\textbf{Another common assumption concerns the slope of the power law decline, taken to be $s= -1$.
This is also not correct, since this slope is a function of the shape of the distribution of the separations of the DNS systems at birth} (item (ii) above). 
The power law regime, which follows the plateaux, sets in at a delay time which depends on the minimum separation of the DNS systems at birth. In the appendix these properties are justified analytically.  \\

We notice that the DDT for merging DNS in \citet{eldridge19} BPS models show similar characteristics, namely an early peak of $\sim$ 30 Myr duration, followed by a power law decline with slope -1. While the results of BPS codes can be ascribed to the combination of the prescriptions adopted in the calculation, in our derivation these characteristics are directly related to basic astrophysical properties, i.e. the mass of neutron stars progenitors, and the distribution of the separations of the DNS systems at birth. This makes it possible to obtain an easy, yet effective, parametrization of DDT. \\ 

Figure \ref{fig_prompt} shows the fraction (\Fp) of systems which merge within 32 Myr for the variety of combination of the parameters considered, all of which appear to  have some impact on this fraction. As the distribution of \ai\ becomes steeper \Fp\ increases, and, for a given $\beta$, the lower the limit on \ai\ the larger the fraction of \prompt\ events. These two parameters appear equally important, with, e.g., the same value of \Fp\ obtained 
with the combinations ($\beta=-1,\ain=0.2 \rsun$) and ($\beta=-2,\ain=0.5\rsun$). The fraction of \prompt\ events also depends on the distributions of the total mass and of the eccentricity of the DNS systems, although to a lesser extent, especially when the distribution of \ai\ has a large abundance of binaries with low separations. We notice that the distributions discussed so far do not consider the possibility of a correlation between the separation \ai\ and the mass of the DNS system which may result from a more effective shrinkage of the more massive systems during the CE phase, similar to the \textit{CLOSE DD} variety of \citet{greggio05} models for SNe Ia. If this were the case, for a given $\beta$, the distribution of the GWR delays would be steeper, and the fraction of prompt events would be higher. We tested this possibility computing the DDT for the cases \ain=0.5 \rsun, $\beta = -1,-3$, $\rho=-0.5,0,1$, and assuming the (arbitrary) relation $\mdn-2 = 1.4/\sqrt{\ai}$. The results are shown on Fig. \ref{fig_prompt} as green crosses. It appears that the effect of adopting a correlation between \mdn\ and \ai\ is modest, and on the same order of that due to a variation of the distribution of the eccentricities at fixed $\beta$ and \ain. \\

It turns out that the currently available data on the cosmic rate of SGRBs do not lead to a strong constraint on the parameters controlling the DDT, with the various models yielding an acceptable representation of the data, within the observational uncertainty. A mild preference for models with a sizable component at relatively long delay times (e.g. $\beta=-1$) is present. More stringent indications can be achieved with a larger database of SGRBs (and kilonova) events. The measurement of the SGRB rates in galaxies with different star formation history, and the correlation with the properties of the parent galaxies, can greatly help in assessing the shape of the DDT, as currently done for SNe Ia \citep[e.g.][]{botticella17}. The association with early type galaxies of some SGRBs, and of the GW170817 event, support a  shallow slope for the DDT of merging DNS. 
However, due to the steep slope of the DDT (e.g. dropping by factor of $10^{3.8}$ from 20 Myr to 10 Gyr), a modest recent star formation activity could give rise to a merging DNS even in a generally old galaxy. 
More precise constraints will come from future observational campaigns aimed at measuring the rate of these events in galaxies of different type.

Notwithstanding the loose constraints on the DDT from the redshift distribution of SGRBs, their local rate, as well as the local rate of kilonovae, allows us to derive a robust indication of the fraction of neutron star progenitors which should follow the evolutionary path leading to 
DNS systems merging within  a Hubble time. This fraction turns out to be $\alpha_{\rm MNS} \sim 1 \%$. 
A number of uncertainties bear upon the determination of \alphamns, including those related to the local rate of kilonovae, the approximations introduced in our procedure to evaluate \kkn, and the possible systematic with age and metallicity of the kilonova production from stellar populations. Nevertheless, had we found a much different value for \kkn, and in turn \alphamns, it would be hard to account for the solar Europium abundance with the standard chemical evolution model.
We point out that in the chemical evolution model we follow not only  the [Eu/Fe] vs. [Fe/H], but also the abundances of the elements produced by massive stars in the same mass range of merging neutron stars. Therefore, also the Fe  and $\alpha$-element abundances are affected by the adopted value of \alphamns.  Our chemical evolution model follows 40 species and it is aimed at reproducing the [X/Fe] vs. [Fe/H] relations together with the solar abundances of all the considered elements.

 With the same approach adopted here \citet{greggio19} find that in order to account for the observed rate of SNe Ia in supernova surveys,
$\sim 3 \%$ of the stars with mass between 2.5 and 9 \msun\ should be found in systems which evolve to the final explosion. Although the details of the evolution of the two kinds of explosive events are different, still \textbf{the observed rates indicate that a few percent of the progenitors should be found in close binaries of the variety which secures the final explosion in a Hubble time}. 
This is a strong constraint to the BPS models and may be used to appraise the input ingredients, for example those which impact on the fraction of systems which are disrupted in the course of the evolution.
\\

With our measurement of $\alpha_{\rm MNS}$ we have computed models for the trend of [Eu/Fe] versus [Fe/H] in the chemical evolution of the Milky Way.  
If merging DNS are the only contributors to the Europium abundance in the Interstellar Medium, a very prompt release of this element is necessary to reproduce the high [Eu/Fe] ratio in low metallicity, halo stars. However even with the DDT which in our models has the highest \textit{prompt} fraction  we cannot match the observed level of the abundance ratio. Resorting to a steeper distribution of the separations (lower $\beta$) or to an even smaller minimum separation (\ain) does not solve the problem because the curves become insensitive to these parameters.
One possible solution consists in assuming some Eu production from CC SNe which would release Europium to the ISM at the same pace as their Iron production. In our models, the data are well matched with a comparable contribution to Eu from CC SNe and from kilonovae. While the discrepancy on the [Eu/Fe] ratio  is more critical for the low metallicity stars, the assumption of a contribution to Europium from CC-SNe leads to a better match over the whole Fe range. Unfortunately, in this case, the Europium production from CC SNe obscures the contribution from kilonovae, so that it is very difficult to discriminate between the various DDTs. 
\\

To conclude, based on a thorough exploration of the possibilities for the distribution of the delay times of merging DNS, the current data on the redshift distribution of SGRBs and on the trend of the Europium to Fe ratio in Milky Way stars leads to the following conclusions:
\begin{itemize}
\item approximately 1 \% of the neutron star progenitors should be found in binary systems which evolve up to the final exploding event;
\item an additional source of Europium is required  to account for the high [Eu/Fe] ratio in the Galactic halo stars, besides the kilonova events. This applies to a lesser extent also to disk stars, so that a contribution from ordinary CC-SNe is favoured;
\item no strong constraint on the DDT can currently be derived from the observations, so that it is possible to match all the currently available data with the same DDT.
\end{itemize}

\section*{Acknowledgements}

FM acknowledges funds from University of Trieste (FRA2016). 
PS wish to thank Marta Molero for the technical support in running some of the chemical models. PS also acknowledges the European Space Agency for co-funding his doctoral project (EXPRO RFP IPL-PSS/JD/190.2016).

\section*{Data Availability} 

Data are available upon request.




\nocite{battistiniCat}
\bibliographystyle{mnras}
\bibliography{MNS_DTD} 



%
%
%
%
\appendix
\section{Analytical derivation of the distribution of the coalescence times}
\begin{figure}
\includegraphics[width=\columnwidth]{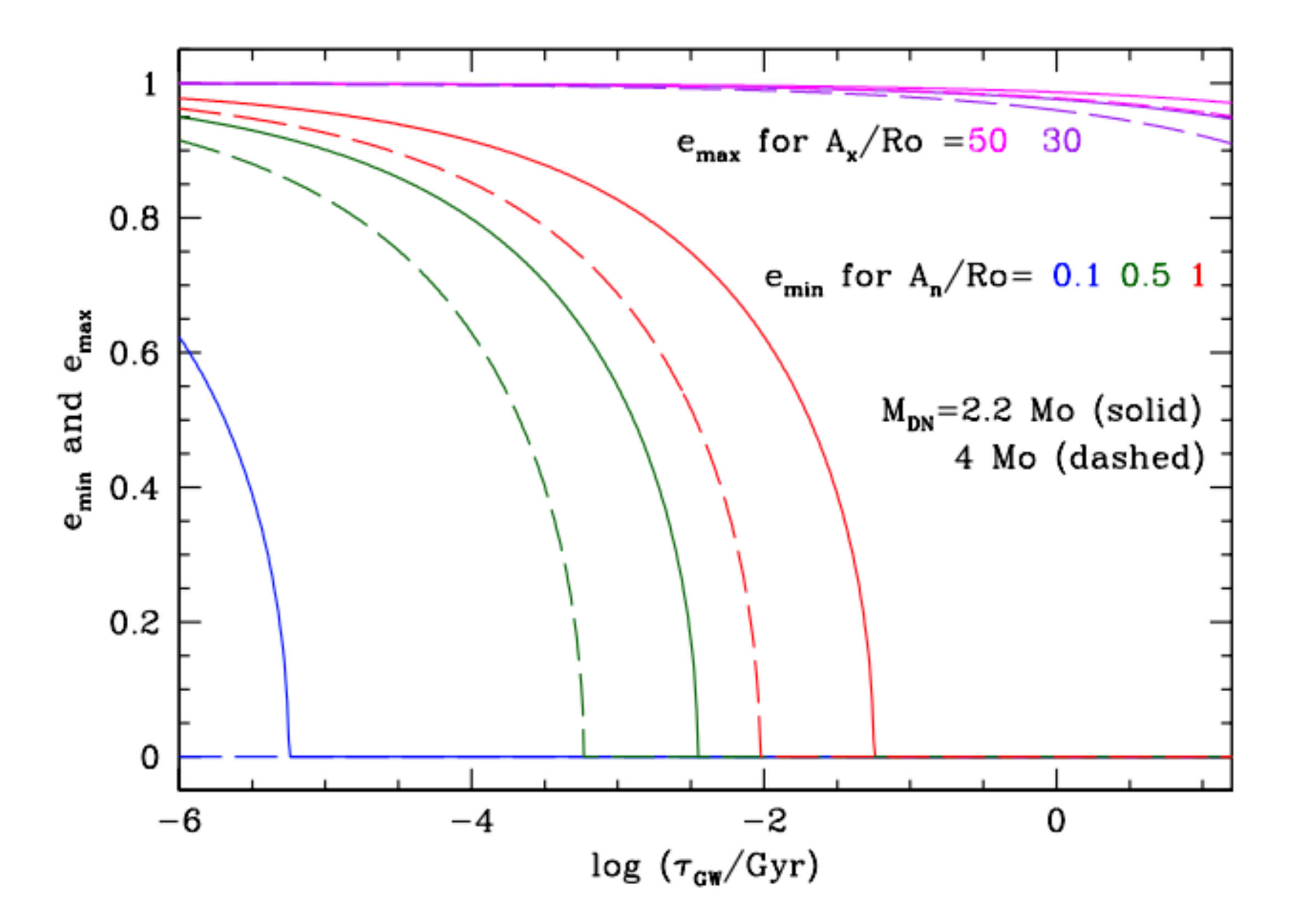}
    \caption{Limits on eccentricity imposed by the limits on the separations.}
    \label{fig_app1}
\end{figure}

\begin{figure*}
\centering
\resizebox{\hsize}{!}{
\includegraphics[angle=0,clip=true]{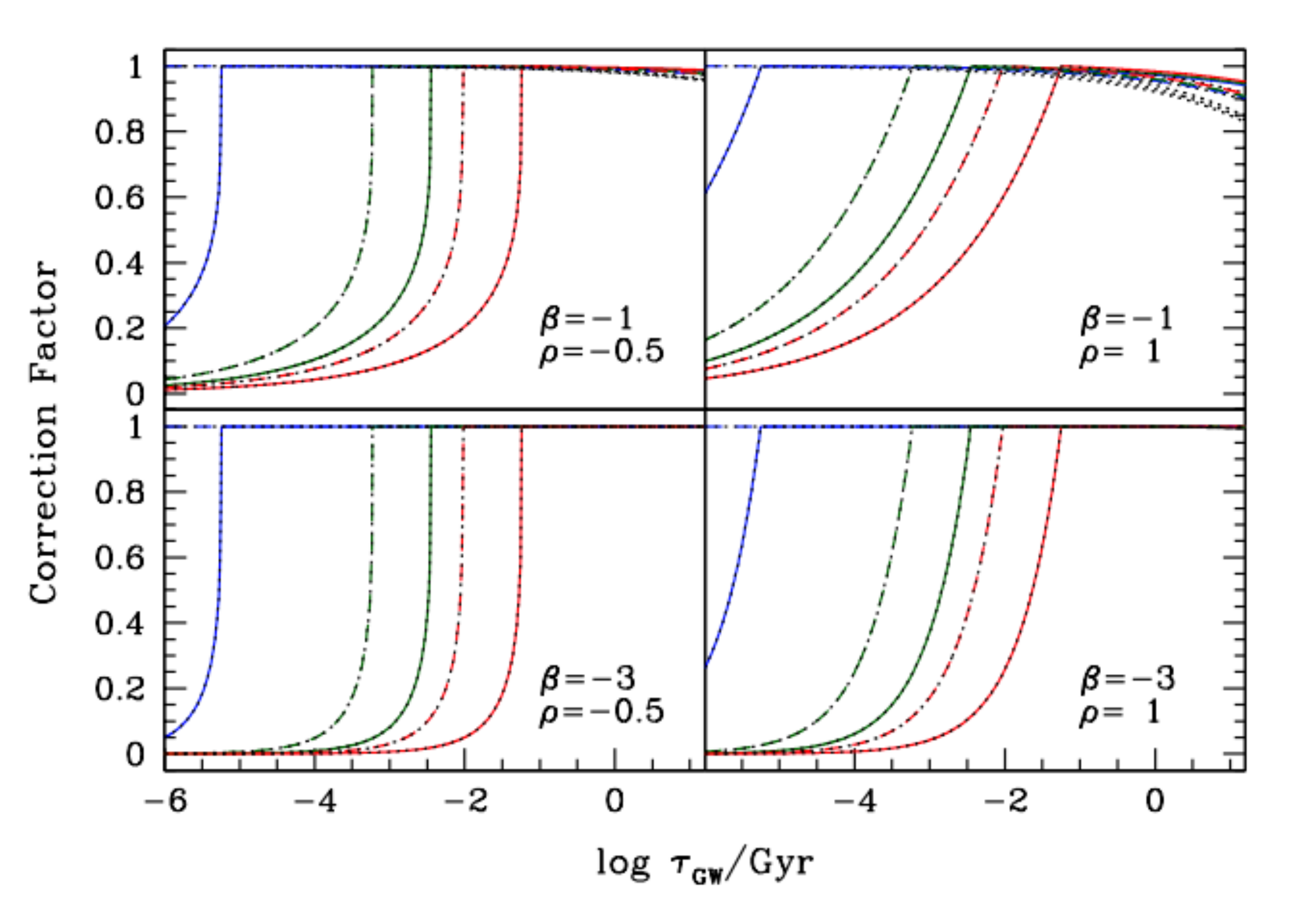}}
\caption{Correction factor to the straight power law as a function of \taugw\ for various options of the parameters $\beta$ and $\rho$, as labelled. Solid lines are computed with \mdn=2.2 \msun, dashed lines with \mdn=4 \msun. Blue, green and red lines show the correction factor respectively for \ain=0.1,0.5 and 1 \rsun and \aix=50 \rsun. Black dotted lines illustrate how the correction factor is modified when \aix=30 \rsun is adopted.}
\label{fig_app2}
\end{figure*}
The time taken by the gravitational waves radiation to bring into contact the two neutron stars members of a binary with total mass, initial separation and eccentricity $(\mdn,A,e)$ can be approximated as in Eq.(\ref{eq_taugw}). In \citet{simonetti19} we derived the distribution of \taugw\ for systems with $e=0$; here we take into account the dependence on the eccentricity, but consider the case of constant $\mdn = \mdnzero$.

The contribution to the number of systems with delay \taugw\ from systems with total mass \mdnzero, separation A and eccentricity $e$ is :

\begin{equation}
{\rm d}n(\taugw,e) = n(e) \times n(\astar)\, {\rm d}e \, {\rm d}\astar,
\label{eq_A1}
\end{equation}

\noindent
where $n(e),n(A)$ are the distribution functions of the eccentricity and of the separations, the latter evaluated at

\begin{equation}
\astar = \Bigl (\frac{\mdnzero^3 \,  \taugw}{0.6 \, (1-e^2)^{3.5}} \Bigr ) ^{0.25} \,\,\, \rsun.
\label{eq_A2}
\end{equation}

\noindent
with \mdnzero\ in solar masses and \taugw\ in Gyr.
Integrating on the eccentricities we get

\begin{equation}
n(\taugw){\rm d}\taugw={\rm d}\taugw \, \int_{0}^{1}  n(e) \times n(\astar) \Bigl |\frac{\partial A}{\partial \taugw} \Bigr | \, {\rm d}e .
\label{eq_A3}
\end{equation}

\noindent
Substituting 

\begin{equation}
\frac{\partial A}{\partial \taugw} \propto \taugw^{-0.75} \times (1-e^2)^{-7/8} 
\label{eq_A4}
\end{equation}

\noindent
eq. (\ref{eq_A3}) becomes

\begin{equation}
  n(\taugw) \propto \taugw^{-0.75} \int_0^1 
  n(e) \times n(\astar) \times (1-e^2)^{-7/8} {\rm d} e .  
  \label{eq_A5}
\end{equation}

\noindent
Adopting $n(e) \propto e^\rho$ and 

\begin{equation}
n(A) \propto
\begin{cases}
0 \,\,\,\,\,\quad  {\rm in} \quad A < \ain \\
A^{\beta} \,\,\, \quad  {\rm in} \quad \ain \leq A \leq \aix\\
0 \,\,\,\,\,\quad  {\rm in} \quad A > \aix
\end{cases}
\label{eq_A6}
\end{equation}

\noindent
we get

\begin{equation}
  n(\taugw) \propto \taugw^{-0.75+0.25\beta} \int_{\emin}^{\emax} 
   \frac{e^\rho}{(1-e^2)^{7(1+\beta)/8}} {\rm d} e   
  \label{eq_A7}
\end{equation}

\noindent
where \emin\ and \emax\ are respectively the minimum and maximum values of the eccentricity which correspond to \ain\ and \aix.
Eq. (\ref{eq_A7}) shows that the distribution of the GWR delays scales  proportionally to a power law with exponent $s=-0.75+0.25 \beta$ modified by a factor, CF(\taugw), which accounts for the limits on the parameter space due to the conditions on the allowed range of separations.
On Fig. \ref{fig_taugw} one can see that at a given \taugw\ a lower limit on $A$ implies  lower limit on $e$ greater than 0, and an upper limit on $A$ implies an upper limit on $e$ smaller than 1. In fact:

\vspace{0.4cm}
$\begin{cases}
\emin^2 = {\rm max} \Bigl [ 0 ; 1-\Bigl (\frac{\taugw \, \mdnzero^3}{0.6 \ain^4} \Bigr )^{2/7} \Bigr ] \\
\emax^2 = {\rm min} \Bigl [ 1 ; 1-\Bigl (\frac{\taugw \, \mdnzero^3}{0.6 \aix^4} \Bigr )^{2/7} \Bigr ] .
\end{cases}$
\vspace{0.4cm}

\noindent
These limits are shown on Fig. \ref{fig_app1} as function of \taugw\ for our adopted minimum and maximum values of \mdn, and various options for the range of separations. If \aix\ is sufficiently large, the upper limit \emax\ is close to 1 for all GWR delays within a Hubble time, irrespectively of \mdn. The effect of the lower limit \ain\ is much more important: it implies a shrinkage of the range of eccentricities providing short GWR delays by an amount which depends on both \ain\ and \mdn.

Figure \ref{fig_app2} shows the correction factor CF(\taugw) normalized to its maximum value, for a few combinations of the parameters.  
\begin{itemize}
    \item The correction factor is very small at short \taugw; it increases with increasing \taugw\ up to a maximum reached at \taugwstar, beyond which it remains constant. 
    \item the value of \taugwstar\ increases with \ain\ increasing, and, for \ain=1 \rsun\, it is of (9.5, 57) Myr respectively for  \mdn=4 and 2,2 \msun;
    \item the correction factor depends on \aix\ only if the distribution of the separation is relatively flat, and even in that case it appears quite mild;
    \item steeper values of $\beta$ and flatter values of $\rho$ lead to a sharper variation of CF(\taugw) approaching \taugwstar. 
\end{itemize}
Eq. \ref{eq_A7} and the correction factor shown on Fig. \ref{fig_app2} have been derived under the assumption that the variables $(A,e)$ are independent. When a correlation is introduced between $A_f/A_i$ and eccentricity, as in the cases shown on Figs \ref{fig_ntg_A3} and \ref{fig_ntg_Arange}, some modifications on the correction factor set in, especially in the vicinity of \taugwstar. 

\bsp	
\label{lastpage}
\end{document}